%% file: notebook.tex
\documentclass[dvipsnames]{article}
\usepackage{fullpage}
\usepackage[square,numbers]{natbib}
\usepackage{color}
\usepackage{amsfonts}
\usepackage{amsmath}
\usepackage{xcolor}
\usepackage{hyperref}
\usepackage{graphicx}
\usepackage{booktabs}
\usepackage{notation}
\usepackage{pandoc}
\usepackage[inline, shortlabels]{enumitem}
\setlist{itemjoin ={,\enspace},itemjoin* = { and\enspace}}

\begin{document}
\title{Overview of the TREC 2021 Fair Ranking Track}
\author{
  Michael D. Ekstrand\\
  \texttt{michaelekstrand@boisestate.edu}
  \and
  Graham McDonald \\
  \texttt{graham.mcdonald@glasgow.ac.uk}
  \and
  Amifa Raj\\
  \texttt{amifaraj@u.boisestate.edu}
  \and
  Isaac Johnson\\
  \texttt{isaac@wikimedia.org}
}

\maketitle

\input{introduction}

\input{task-definition}

\input{data}

\input{evaluation}
\input{results}
\input{limitations}
\bibliography{proposal}
\appendix
\input{nb-alignments}

\end{document}

%% file: introduction.tex
\section{Introduction}
The TREC Fair Ranking Track aims to provide a platform for participants to develop and evaluate novel retrieval algorithms that can provide a fair exposure to a mixture of demographics or attributes, such as ethnicity, that are represented by relevant documents in response to a search query. For example, particular demographics or attributes can be represented by the documents' topical content or authors.

The 2021 Fair Ranking Track adopted a resource allocation task. The task focused on supporting Wikipedia editors who are looking to improve the encyclopedia's coverage of topics under the purview of a WikiProject.\footnote{\url{https://en.wikipedia.org/wiki/WikiProject}} WikiProject coordinators and/or Wikipedia editors search for Wikipedia documents that are in need of editing to improve the quality of the article. The 2021 Fair Ranking track aimed to ensure that documents that are about, or somehow represent, certain protected characteristics receive a fair exposure to the Wikipedia editors, so that the documents have an fair opportunity of being improved and, therefore, be well-represented in Wikipedia. The under-representation of particular protected characteristics in Wikipedia can result in systematic biases that can have a negative human, social, and economic impact, particularly for disadvantaged or protected societal groups~\cite{pedreshi2008discrimination,redi2020taxonomy}.

%% file: task-definition.tex
\section{Task Definition}\label{sec:task-definition}
The 2021 Fair Ranking Track used an \emph{ad hoc} retrieval protocol. Participants were provided with a corpus of documents (a subset of the English language Wikipedia) and a set of queries. A query was of the form of a short list of search terms that represent a WikiProject. Each document in the corpus was relevant to zero to many WikiProjects and associated with zero to many fairness categories.   

There were two tasks in the 2021 Fair Ranking Track. In each of the tasks, for a given query, participants were to produce document rankings that are: 

\begin{enumerate}
\item Relevant to a particular WikiProject.
\item Provide a fair exposure to articles that are associated to particular protected attributes.
\end{enumerate}

The tasks shared a topic set, the corpus, the basic problem structure and the fairness objective. However, they differed in their target user persona, system output (static ranking vs.\ sequences of rankings) and evaluation metrics. The common problem setup was as follows:

\begin{itemize}
    \item \textbf{Queries} were provided by the organizers and derived from the topics of existing or hypothetical WikiProjects.
    \item \textbf{Documents} were Wikipedia articles that may or may not be relevant to any particular WikiProject that is represented by a query.
    \item \textbf{Rankings} were ranked lists of articles for editors to consider working on.
    \item \textbf{Fairness} of exposure was achieved with respect to the \textbf{geographic location} of the articles (geographic location annotations were provided). For the evaluation topics, in addition to geographic fairness, to the extent that biographical articles are relevant to the topic, the rankings should have also been fair with respect to an undisclosed \textbf{demographic attribute} of the people that the biographies cover, which was gender.
\end{itemize}

\subsection{Task 1: WikiProject Coordinators}\label{subsec:task1}
The first task focused on WikiProject coordinators as users of the search system; their goal is to search for relevant articles and produce a ranked list of articles needing work that other editors can then consult when looking for work to do.\\

\noindent \textbf{Output}: The output for this task was a \textbf{single ranking per query}, consisting of \textbf{1000 articles}.\\

Evaluation was a multi-objective assessment of rankings by the following two criteria:

\begin{itemize}
    \item Relevance to a WikiProject topic. Relevance assessments were provided for articles for the training queries derived from existing Wikipedia data; evaluation query relevance were assessed by NIST assessors. Ranking relevance was computed with nDCG, using binary relevance and logarithmic decay.
    \item Fairness with respect to the exposure of different fairness categories in the articles returned in response to a query.
\end{itemize}

Section \ref{sec:eval-task1} contains details on the evaluation metrics.

\subsection{Task 2: Wikipedia Editors}\label{subsec:task2}

The second task focused on individual Wikipedia editors looking for work associated with a project.
The conceptual model is that rather than maintaining a fixed work list as in Task 1, a WikiProject coordinator would create a saved search, and when an editor looks for work they re-run the search.
This means that different editors may receive different rankings for the same query, and differences in these rankings may be leveraged for providing fairness.\\

\noindent \textbf{Output}: The output of this task is \textbf{100 rankings per query}, each consisting of \textbf{50 articles}. \\

Evaluation was a multi-objective assessment of rankings by the following three criteria:

\begin{itemize}
    \item Relevance to a WikiProject topic. Relevance assessments were provided for articles for the training queries derived from existing Wikipedia data; evaluation query relevance was assessed by NIST assessors. Ranking relevance was computed with nDCG.
    \item Work needed on the article (articles needing more work preferred). We provided the output of an article quality assessment tool for each article in the corpus; for the purposes of this track, we assumed lower-quality articles need more work. 
    \item Fairness with respect to the exposure of different fairness categories in the articles returned in response to a query.
\end{itemize}

The goal of this task was \textit{not} to be fair to work-needed levels; rather, we consider work-needed and topical relevance to be two components of a multi-objective notion of relevance, so that between two documents with the same topical relevance, the one with more work needed is more relevant to the query in the context of looking for articles to improve.

This task used \emph{expected exposure} to compare the exposure article subjects receive in result rankings to the \emph{ideal} (or \emph{target}) \emph{exposure} they would receive based on their relevance and work-needed \cite{diaz2020evaluating}.
This addresses fundamental limits in the ability to provide fair exposure in a single ranking by examining the exposure over multiple rankings.

For each query, participants provided 100 rankings, which we considered to be samples from the distribution realized by a stochastic ranking policy (given a query $\query$, a distribution $\pi_\query$ over truncated permutations of the documents).
Note that this is how we interpret the queries, but it did not mean that a stochastic policy is how the system should have been implemented --- other implementation designs were certainly possible.
The objective was to provide equitable exposure to documents of comparable relevance and work-needed, aggregated by protected attribute. Section~\ref{sec:eval-task2} has details on the evaluation metrics.

%% file: data.tex
\section{Data}
This section provides details of the format of the test collection, topics and ground truth. Further details about data generation and limitations can be found in Section~\ref{sec:limitations}.

\subsection{Obtaining the Data}

The corpus and query data set is distributed via Globus, and can be obtained in two ways. First, it can be obtained via Globus, from our repository at \url{https://boi.st/TREC2021Globus}. From this site, you can log in using your institution's Globus account or your own Google account, and synchronize it to your local Globus install or download it with Globus Connect Personal.\footnote{\url{https://www.globus.org/globus-connect-personal}}  This method has robust support for restarting downloads and dealing with intermittent connections. Second, it can be downloaded directly via HTTP from:\\ \url{https://data.boisestate.edu/library/Ekstrand-2021/TRECFairRanking2021/}. 

The runs and evaluation qrels will be made available in the ordinary TREC archives.

\subsection{Corpus}\label{subsec:corpus}

The corpus consisted of articles from English Wikipedia.
We removed all redirect articles, but left the wikitext (markup Wikipedia uses to describe formatting) intact.
This was provided as a JSON file, with one record per line, and compressed with gzip (\texttt{trec\_corpus.json.gz}). Each record contains the following fields:

\begin{description}
    \item[id] The unique numeric Wikipedia article identifier.
    \item[title] The article title.
    \item[url] The article URL, to comply with Wikipedia licensing attribution requirements.
    \item[text] The full article text.
\end{description}

The contents of this corpus were prepared in accordance with, and licensed under, the CC BY-SA 3.0 license.\footnote{\url{https://creativecommons.org/licenses/by-sa/3.0/}} The raw Wikipedia dump files used to produce this corpus are available in the \texttt{source} directory; this is primarily for archival purposes, because Wikipedia does not publish dumps indefinitely.

\subsection{Topics}\label{subsec:topics}

Each of the track's training topics is based on a single Wikiproject. The topic is also GZIP-compressed JSON lines (file \texttt{trec\_topics.json.gz}), with each record containing:

\begin{description}
    \item[id] A query identifier (int)
    \item[title] The Wikiproject title (string)
    \item[keywords] A collection of search keywords forming the query text (list of str)
    \item[scope] A textual description of the project scope, from its project page (string)
    \item[homepage] The URL for the Wikiproject. This is provided for attribution and not expected to be used by your system as it will not be present in the evaluation data (string)
    \item[rel\_docs] A list of the page IDs of relevant pages (list of int)
\end{description}

The keywords are the primary query text. The scope is there to provide some additional context and potentially support techniques for refining system queries.

In addition to topical relevance, for Task 2: Wikipedia Editors (Section~\ref{subsec:task2}), participants were also expected to return relevant documents that need more editing work done more highly than relevant documents that need less work done.

\subsection{Annotations}
NIST assessors annotated the retrieved documents with binary relevance score for given topics. We provided additional options like \textit{unassessable} and  \textit{skip} if the document-topic pair is difficult to assess or the assessor is not familiar with the topic. The annotations are incomplete, for reasons including:

\begin{itemize}
    \item Task 2 requires sequence of rankings which results a large number of dataset, thus it was not possible to annotate all the retrieved documents.
    \item Some documents were not complete and did not have enough information to match with the topic.
\end{itemize}

We obtained assessments through tiered pooling, with the goal of having assessments for a coherent subset of rankings that are as complete as possible.  We have assessments for the following tiers:

\begin{itemize}
    \item The first 20 items of all rankings for Task 1 (all queries).
    \item The first 5 items of the first 25 rankings from every submission to Task 2 (about 75\% of the queries).
\end{itemize}

Details are included with the annotations and metric code.

\subsection{Metadata and Fairness Categories}\label{subsec:fairnesscat}

For training data, participants were provided with a geographical fairness ground truth. For the evaluation data, submitted systems were evaluated on how fair their rankings are to the geographical fairness category and an undisclosed personal demographic attribute (gender).

We also provided a simple Wikimedia quality score (a float between 0 and 1 where 0 is no content on the page and 1 is high quality) for optimizing for work-needed in Task 2. Work-needed was operationalized as the reverse---i.e. 1 minus this quality score.  The discretized quality scores were used as work-needed for final system evaluation.

This data was provided together in a metadata file (\texttt{trec\_metadata.json.gz}), in which each line is the metadata for one article represented as a JSON record with the following keys:

\begin{description}
    \item[page\_id] Unique page identifier (int)
    \item[quality\_score] Continuous measure of article quality with 0 representing low quality and 1 representing high quality (float in range $[0,1]$)
    \item[quality\_score\_disc] Discrete quality score in which the quality score is mapped to six ordinal categories from low to high: Stub, Start, C, B, GA, FA (string)
    \item[geographic\_locations] Continents that are associated with the article topic. Zero or many of: Africa, Antarctica, Asia, Europe, Latin America and the Caribbean, Northern America, Oceania (list of string)
    \item[gender] For articles with a gender, the gender of the article's subject, obtained from WikiData.
\end{description}

\subsection{Output}\label{subsec:output}

For \textbf{Task 1}, participants outputted results in rank order in a tab-separated file with two columns:

\begin{description}
    \item[id] The query ID for the topic
    \item[page\_id] ID for the recommended article
\end{description}

\noindent For \textbf{Task 2}, this file had 3 columns, to account for repeated rankings per query:

\begin{description}
    \item[id] Query ID
    \item[rep\_number] Repeat Number (1-100)
    \item[page\_id] ID for the recommended article
\end{description}

%% file: evaluation.tex
\section{Evaluation Metrics}

Each task was evaluated with its own metric designed for that task setting. The goal of these metrics was to measure the extent to which a system (1) exposed relevant documents, and (2) exposed those documents in a way that is fair to article topic groups, defined by location (continent) and (when relevant) the gender of the article's subject.

This faces a problem in that Wikipedia itself has well-documented biases: if we target the current group distribution within Wikipedia, we will reward systems that simply reproduce Wikipedia's existing biases instead of promoting social equity. However, if we simply target equal exposure for groups, we would ignore potential real disparities in topical relevance. Due to the biases in Wikipedia's coverage, and the inability to retrieve documents that don't exist to fill in coverage gaps, there is not good empirical data on what the distribution for any particular topic \textit{should} be if systemic biases did not exist in either Wikipedia or society (the ``world as it could and should be'' \citep{mitchell:fairness}). Therefore, in this track we adopted a compromise: we \textbf{averaged} the empirical distribution of groups among relevant documents with the world population (for location) or equality (for gender) to derive the target group distribution.

Code to implement the metrics is found at \url{https://github.com/fair-trec/trec2021-fair-public}.

\subsection{Preliminaries}

The tasks were to retrieve documents $\doc$ from a corpus $\Corpus$ that are relevant to a query $\query$.  $\relvec{\query} \in [0,1]^{|\Corpus|}$ is a vector of relevance judgements for query $q$.
We denote a ranked list by $\ranking$; $\ranking_i$ is the document at position $i$ (starting from 1), and $\rankInv_\doc$ is the rank of document $\doc$.
For Task 1, each system returned a single ranked list; for Task 2, it returned a sequence of rankings $\rankSeq$.

We represented the group alignment of a document $\doc$ with an \textit{alignment vector} $\avec{\doc} \in [0,1]^{|\groups|}$. $\dal{\doc}{\group}$ is document $\doc$'s alignment with group $\group$.  $\amat \in [0,1]^{|\Corpus| \times |\groups|}$ is the alignment matrix for all documents. $\avec{\world}$ denotes the distribution of the world.\footnote{Obtained from \url{https://en.wikipedia.org/wiki/List_of_continents_and_continental_subregions_by_population}}

We considered fairness with respect to two group sets, $\groups_\geo$ and $\groups_\gender$.
We operationalized this intersectional objective by letting $\groups = \groups_\geo \times \groups_\gender$, the Cartesian product of the two group sets.
Further, alignment under either group set may be unknown; we represented this case by treating ``unknown'' as its own group ($\group_?$) in each set.
In the product set, a document's alignment may be unknown for either or both groups.

In all metrics, we use \textbf{log discounting} to compute attention weights:

$$\weight_i = \frac{1}{\log_2 \max (i, 2)}$$

Task 2 also considered the work each document needs, represented by $\workDoc \in \{1,2,3,4\}$.

\subsection{Task 1: WikiProject Coordinators (Single Rankings)}
\label{sec:eval-task1}

For the single-ranking Task 1, we adopted attention-weighted rank fairness (AWRF), first described by \citet{sapiezynski2019quantifying} and named by \citet{Raj2020-om}. AWRF computes a vector $\dlist$ of the cumulated exposure a list gives to each group, and a target vector $\dtarget$; we then compared these with the Jenson-Shannon divergence:

\begin{align}
\dlist' & = \sum_i \weight_i \avec{\ranking_i} & \text{cumulated attention} \nonumber\\
\dlist & = \frac{\dlist'}{\|\dlist'\|_1} & \text{normalize to a distribution} \nonumber \\
\dtarget & = \frac{1}{2}\left( \amat^\transpose \relvec{\query} + \avec{\world} \right) \nonumber \\
\AWRF(\ranking) & = 1 - \Djs{\dlist}{\dtarget}
\end{align}

For Task 1, we ignored documents that are fully unknown for the purposes of computing $\dlist$ and $\dtarget$; they do not contribute exposure to any group.

The resulting metric is in the range $[0,1]$, with 1 representing a maximally-fair ranking (the distance from the target distribution is minimized). We combined it with an ordinary nDCG metric for utility:

\begin{align}
    \nDCG(\ranking) = \frac{\sum_i \weight_i \rel{\query}{\doc}}{\mathrm{ideal}} \\
    \MOne(\ranking) = \AWRF(\ranking) \times \nDCG(\ranking)
\end{align}

To score well on the final metric $\MOne$, a run must be \textbf{both} accurate and fair.

\subsection{Task 2: Wikipedia Editors (Multiple Rankings)}
\label{sec:eval-task2}

For Task 2, we used Expected Exposure \citep{diaz2020evaluating} to compare the exposure each group receives in the sequence of rankings to the exposure it would receive in a sequence of rankings drawn from an \textit{ideal policy} with the following properties:

\begin{itemize}
    \item Relevant documents come before irrelevant documents
    \item Relevant documents are sorted in nonincreasing order of work needed
    \item Within each work-needed bin of relevant documents, group exposure is fairly distributed according to the average of the distribution of relevant documents and the distribution of global population (the same average target as before).
\end{itemize}

We have encountered some confusion about whether this task is requiring fairness towards work-needed; as we have designed the metric, work-needed is considered to be a part of (graded) relevance: a document is more relevant if it is relevant to the topic and needs significant work.  In the Expected Exposure framework, this combined relevance is used to derive the target policies.

To apply expected exposure, we first define the exposure $\docExp$ a document $\doc$ receives in sequence $\rankSeq$:

\begin{equation}
    \docExp = \frac{1}{|\rankSeq|} \sum_{\ranking \in \rankSeq} w_{\rankInv_\doc}
\end{equation}

This forms an exposure vector $\expVec \in \mathbb{R}^{|\Corpus|}$.  It is aggregated into a group exposure vector $\groupExp$, including ``unknown'' as a group:

\begin{equation}
    \groupExp = \amat^\transpose \expVec
\end{equation}

Our implementation rearranges the mean and aggregate operations, but the result is mathematically equivalent.

We then compare these system exposures with the target exposures $\tgtExpVec$ for each query.  This starts with the per-document ideal exposure; if $m_\work$ is the number of relevant documents with work-needed level $\work \in \{1, 2, 3, 4\}$, then according to \citet{diaz2020evaluating} the ideal exposure for document $\doc$ is computed as:

\begin{equation}
    \targetExposure_\doc = \frac{1}{m_{\workDoc}} \sum_{i = m_{>\workDoc} + 1}^{m_{\ge \workDoc}} \weight_i
\end{equation}

We use this to compute the non-averaged target distribution $\tilde\groupExp^*$:

\begin{equation}
    \tilde\groupExp^* = \amat^\transpose \tgtExpVec
\end{equation}

Since we include ``unknown'' as a group, we have a challenge with computing the target distribution by averaging the empirical distribution of relevant documents and the global population --- global population does not provide any information on the proportion of relevant articles for which the fairness attributes are relevant.  Our solution, therefore, is to average the distribution of \emph{known-group} documents with the world population, and re-normalize so the final distribution is a probability distribution, but derive the proportion of known- to unknown-group documents entirely from the empirical distribution of relevant documents.  Extended to handle partially-unknown documents, this procedure proceeds as follows:

\begin{itemize}
    \item Average the distribution of fully-known documents (both gender and location are known) with the global intersectional population (global population by location and equality by gender).
    \item Average the distribution of documents with unknown location but known gender with the equality gender distribution.
    \item Average the distribution of documents with unknown gender but known location with the world population.
\end{itemize}

The result is the target group exposure $\groupExp^*$.  We use this to measure the \textbf{expected exposure loss}:

\begin{align}
    \MTwo(\rankSeq_\query) & = \| \groupExp - \groupExp^* \|_2 \\
    & = \groupExp \cdot \groupExp - 2\groupExp \cdot \groupExp^* + \groupExp^* \cdot \groupExp^* \nonumber \\
    \operatorname{EE-D}(\rankSeq_\query) & = \groupExp^* \cdot \groupExp^* \\
    \operatorname{EE-R}(\rankSeq_\query) & = \groupExp \cdot \groupExp^*
\end{align}

Lower $\MTwo$ is better.  It decomposes into two submetrics, the \textbf{expected exposure disparity} (EE-D) that measures overall inequality in exposure independent of relevance, for which lower is better; and the \textbf{expected exposure relevance} (EE-L) that measures exposure/relevance alignment, for which higher is better \citep{diaz2020evaluating}.

%% file: results.tex
\section{Results}\label{sec:results}
\begin{table}[tb]
    \centering
    \input{figures/task1-runs}
    \caption{Task 1 runs. Higher score is better (for all metrics).}
    \label{tab:task1:runs}
\end{table}

\begin{figure}[tb]
    \centering
    \includegraphics[width=0.5\textwidth]{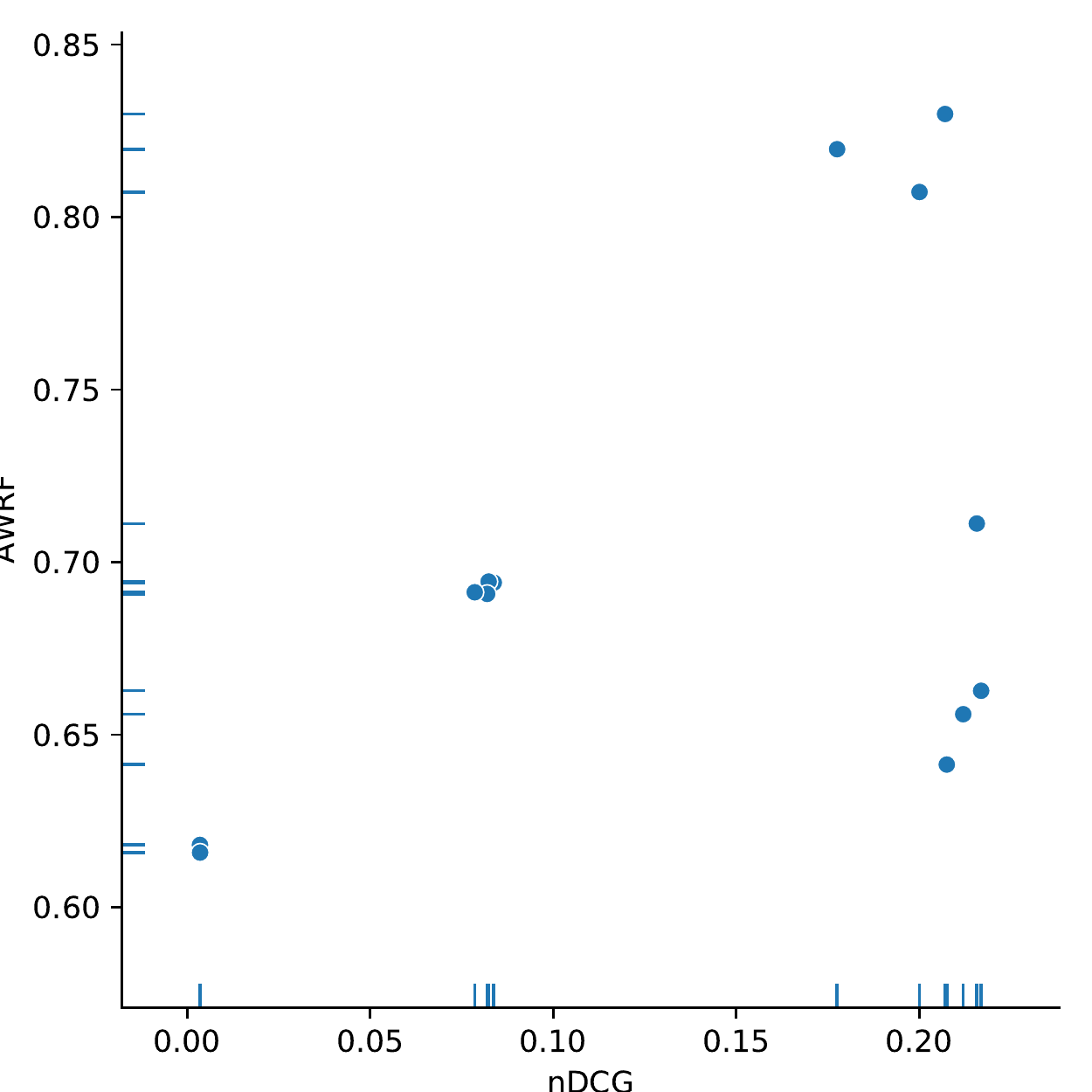}
    \caption{Task 1 submissions by individual component metrics (NDCG and AWRF).  Higher values are better for both metrics.}
    \label{fig:task1:ndcg-awrf}
\end{figure}

This year four different teams submitted a total of 24 runs. All four teams participated in Task 1: Single Rankings (13 runs total), while only three of the four groups participated in Task 2: Multiple Rankings (11 runs total).

\subsection{Task 1: WikiProject Coordinators (Single Rankings)}
Approaches for Task 1 included:
\begin{itemize}
    \item RoBERTa model to compute embeddings for text fields.
    \item A filtering approach to select top ranked documents from either competing rankers or the union of rankers. 
    \item BM25 ranking from pyserini and re-ranked using MMR implicit diversification (without explicit fairness groups). Lambda varied between runs.
    \item BM25 initial ranking with iterative reranking using fairness calculations to select documents to add to the ranking.
    \item Relevance ranking using Terrier plus a fairness component that aims to be fair to both the geographic location attribute and an inferred demographic attribute through tailored diversification plus data fusion.
    \item Optimisation to consider a protected group's distribution in the background collection and the total predicted relevance of the group in the candidate results set.
    \item Allocating positions in the generated ranking to a protected group proportionally with respect to the total relevance score of the group within the candidate results set. 
    \item Relevance-only approaches.
\end{itemize}

Table~\ref{tab:task1:runs} shows the submitted systems ranked by the official Task 1 metric $\MOne$ and its component parts nDCG and AWRF. Figure~\ref{fig:task1:ndcg-awrf} plots the runs with the component metrics on the $x$ and $y$ axes. Notably, each of the approaches from a participating team are clustered in terms of the component metrics and the official $\MOne$ metric.

\subsection{Task 2: Wikipedia Editors (Multiple Rankings)}\label{sub:t2results}
Approaches for Task 2 included:
\begin{itemize}
    \item A randomized method with BERT and a two-staged Plackett-Luce sampling where relevance scores are combined with work needed.
    \item An iterative approach that uses RoBERTa and computes a score for each of the top-K documents in the current state, based on the expected exposure of each group so far and the original estimated relevance score, integrating an article's quality score.
    \item BM25 plus re-ranking iteratively selecting documents by combining relevance, fairness and quality scores.
    \item Relevance ranking using Terrier plus a fairness component that aims to be fair to both the geographic location attribute and an inferred demographic attribute through tailored diversification plus data fusion to prioritise highly relevant documents while matching the distributions of the protected groups in the generated ranking to their distributions in the background population.
    \item Minimising the predicted divergence, or skew, in the distributions of the protected groups over all of the rankings within a sequence, compared to the background population.
    \item Minimising the disparity between a group's expected and actual exposures and learning the importance of the group relevance and background distributions.
    \item Relevance-only ranking.
\end{itemize}

\begin{figure}[tb]
    \centering
    \includegraphics[width=0.5\textwidth]{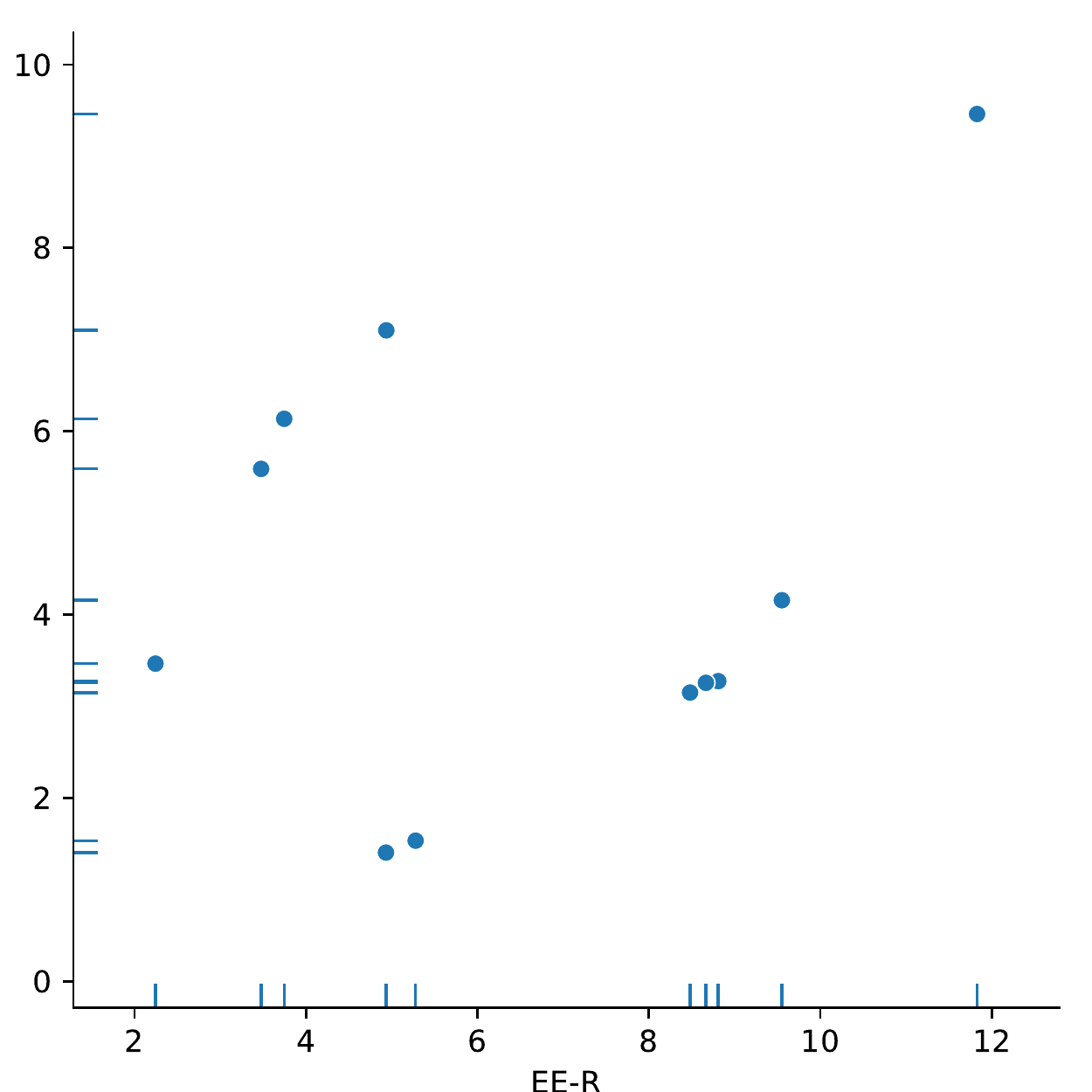}
    \caption{Task 2 submissions by expected exposure subcomponents. Lower EE-D is better; higher EE-R is better.}
    \label{fig:task2:eed-eer}
\end{figure}

Table~\ref{tab:task2:runs} shows the submitted systems ranked by the official Task 2 metric EE-L and its component parts EE-D and EE-R. Figure~\ref{fig:task2:eed-eer} plots the runs with the component metrics on the $x$ and $y$ axes. Overall, the submitted systems generally performed better for one of the component metrics than they did for the other. There are, however, a cluster of four points in Figure~\ref{fig:task2:eed-eer} that make headway in the trade-off between EE-D and EE-L.

\begin{table}[tb]
    \centering
    \input{figures/task2-runs}
    \caption{Task 2 runs. Lower EE-L is better.}
    \label{tab:task2:runs}
\end{table}

%% file: figures/task1-runs.tex
\begin{tabular}{lrrrl}
\toprule
{} &   nDCG &   AWRF &  Score &          95\% CI \\
\midrule
\textbf{UoGTrDExpDisT1 } & 0.2071 & 0.8299 & 0.1761 &  (0.145, 0.212) \\
\textbf{UoGTrDRelDiT1  } & 0.2001 & 0.8072 & 0.1639 &  (0.138, 0.193) \\
\textbf{UoGTrDivPropT1 } & 0.2157 & 0.7112 & 0.1532 &  (0.128, 0.184) \\
\textbf{UoGTrDExpDisLT1} & 0.1776 & 0.8197 & 0.1459 &  (0.122, 0.173) \\
\textbf{RUN1           } & 0.2169 & 0.6627 & 0.1425 &  (0.119, 0.172) \\
\textbf{UoGTrRelT1     } & 0.2120 & 0.6559 & 0.1373 &  (0.113, 0.165) \\
\textbf{RMITRet        } & 0.2075 & 0.6413 & 0.1317 &  (0.110, 0.159) \\
\textbf{1step\_pair     } & 0.0838 & 0.6940 & 0.0648 &  (0.046, 0.090) \\
\textbf{2step\_pair     } & 0.0824 & 0.6943 & 0.0638 &  (0.045, 0.089) \\
\textbf{1step\_pair\_list} & 0.0820 & 0.6908 & 0.0623 &  (0.045, 0.085) \\
\textbf{2step\_pair\_list} & 0.0786 & 0.6912 & 0.0607 &  (0.044, 0.083) \\
\textbf{RMITRetRerank\_1} & 0.0035 & 0.6180 & 0.0026 &  (0.001, 0.009) \\
\textbf{RMITRetRerank\_2} & 0.0035 & 0.6158 & 0.0026 &  (0.001, 0.009) \\
\bottomrule
\end{tabular}

%% file: figures/task2-runs.tex
\begin{tabular}{lrrrc}
\toprule
{} &    EE-R &   EE-D &    EE-L &       EE-L 95\% CI \\
\midrule
\textbf{RUN\_task2      } &  9.5508 & 4.1557 & 14.9007 &  (12.303, 19.946) \\
\textbf{pl\_control\_0.6 } &  8.8091 & 3.2733 & 15.5017 &  (12.552, 20.477) \\
\textbf{UoGTrRelT2     } & 11.8281 & 9.4609 & 15.6514 &  (13.057, 20.148) \\
\textbf{pl\_control\_0.8 } &  8.6654 & 3.2550 & 15.7708 &  (12.746, 21.251) \\
\textbf{pl\_control\_0.92} &  8.4802 & 3.1486 & 16.0348 &  (12.820, 21.158) \\
\textbf{PL\_IRLab\_07    } &  5.2790 & 1.5327 & 20.8213 &  (16.283, 28.089) \\
\textbf{PL\_IRLab\_05    } &  4.9331 & 1.4029 & 21.3832 &  (16.579, 28.293) \\
\textbf{UoGTrDivPropT2 } &  4.9372 & 7.1005 & 27.0726 &  (21.098, 35.870) \\
\textbf{UoGTrDRelDiT2  } &  3.4770 & 5.5891 & 28.4816 &  (22.366, 37.739) \\
\textbf{UoGTrDExpDisT2 } &  3.7459 & 6.1356 & 28.4903 &  (22.571, 37.548) \\
\textbf{UoGTrLambT2    } &  2.2447 & 3.4644 & 28.8216 &  (22.799, 37.718) \\
\bottomrule
\end{tabular}

%% file: limitations.tex
\label{sec:limitations}
\section{Limitations}

The data and metrics in this task address a few specific types of unfairness, and do so partially. This is fundamentally true of any fairness intervention, and does not in any way diminish the value of the effort --- it is impossible for any data set, task definition, or metric to fully capture fairness in a universal way, and all data and analyses have limitations.

Some of the limitations of the data and task include:

\begin{itemize}
    \item \textbf{Fairness criteria}
    \begin{itemize}
        \item \textbf{Geography}: For each Wikipedia article, we ascertained which, if any, continents are relevant to the content.\footnote{Code: \url{https://github.com/geohci/wiki-region-groundtruth/blob/main/wiki-region-data.ipynb}} This was determined by directly looking up several community-maintained (Wikidata) structured data statements about the article. These properties were checked for the presence of countries, which were then mapped to continents via the United Nation's geoscheme.\footnote{\url{https://en.wikipedia.org/wiki/United_Nations_geoscheme}} While this data must meet Wikidata's verifiability guidelines,\footnote{\url{https://www.wikidata.org/wiki/Wikidata:Verifiability}} it does suffer from varying levels of incompleteness. For example, only 73\% of people on Wikidata have a country of citizenship property.\footnote{\url{https://humaniki.wmcloud.org/gender-by-country}} Furthermore, structured data is itself limited---e.g., country of citizenship does not appropriately capture people who are considered stateless though these people may have many strong ties to a country. It is not easy to evaluate whether this data is missing at random or biased against certain regions of the world. Care should be taken when interpreting the absence of associated continents in the data. Further details can be found in the code repository.\footnote{\url{https://github.com/geohci/wiki-region-groundtruth}}
        
        \item \textbf{Gender}: For each Wikipedia article, we also ascertained whether it is a biography, and, if so, which gender identity can be associated with the person it is about.\footnote{Code: \url{https://github.com/geohci/miscellaneous-wikimedia/blob/master/wikidata-properties-spark/wikidata_gender_information.ipynb}} This data is also directly determined via Wikidata based on the instance-of property indicating the article is about a human (P31:Q5 in Wikidata terms) and then collecting the value associated with the sex-or-gender property (P21). Coverage here is much higher at 99.98\% of biographies on Wikipedia having associated gender data on Wikidata.
        
        Assigning gender identities to people is not a process without errors, biases, and ethical concerns. Since we are using it to calculate aggregate statistics, we judged it to be less problematic than it would be if we were making decisions about individuals.  The process for assigning gender is subject to some community-defined technical limitations\footnote{\url{https://www.wikidata.org/wiki/Property_talk:P21\#Documentation}} and the Wikidata policy on living people\footnote{\url{https://www.wikidata.org/wiki/Wikidata:Living_people}}. While a separate project, English Wikipedia's policies on gender identity\footnote{\url{https://en.wikipedia.org/wiki/Wikipedia:Manual_of_Style/Gender_identity}} likely inform how many editors handle gender; in particular, this policy explicitly favors the most recent reliably-sourced \textit{self-identification} for gender, so misgendering a biography subject is a violation of Wikipedia policy; there may be erroneous data, but such data seems to be a violation of policy instead of a policy decision. Wikidata:WikiProject LGBT has documented some clear limitations of gender data on Wikidata and a list of further discussions and considerations.\footnote{\url{https://www.wikidata.org/wiki/Wikidata:WikiProject_LGBT/gender}}
        
        In our analysis (see Appendix A), we handle nonbinary gender identities by using 4 gender categories: unknown, male, female, and third.
        
        We advise great care when working with the gender data, particularly outside the immediate context of the TREC task (either its original instance or using the data to evaluate comparable systems).
    \end{itemize}
    \item \textbf{Relevance Criteria}
    \begin{itemize}
        \item \textbf{WikiProject Relevance}: For the training queries, relevance was obtained from page lists for existing WikiProjects. While WikiProjects have broad coverage of English Wikipedia and we selected for WikiProjects that had tagged new articles in the recent months in the training data as a proxy for activity, it is certain that almost all WikiProjects are incomplete in tagging relevant content (itself a strong motivation for this task). While it is not easy to measure just how incomplete they are, it should not be assumed that content that has not been tagged as relevant to a WikiProject in the training data is indeed irrelevant.\footnote{Current Wikiproject tags were extracted from the database tables maintained by the PageAssessments extension: \url{https://www.mediawiki.org/wiki/Extension:PageAssessments}}
        Evaluation query relevance was assessed by NIST assessors, but the large sets of relevant documents and limited budget for working through the pool mean these lists are also incomplete.
    
        \item \textbf{Work-needed}: Our proxy for work-needed is a coarse proxy. It is based on just a few simple features (page length, sections, images, and references) and does not reflect the nuances of the work needed to craft a top-quality Wikipedia article.\footnote{For further details, see: \url{https://meta.wikimedia.org/wiki/Research:Prioritization_of_Wikipedia_Articles/Language-Agnostic_Quality\#V1}} A fully-fledged system for supporting Wikiprojects would also include a more nuanced approach to understanding the work needed for each article and how to appropriately allocate this work.
    \end{itemize}
    \item \textbf{Task Definition}
    \begin{itemize}
        \item \textbf{Existing Article Bias}: The task is limited to topics for which English Wikipedia already has articles. These tasks are not able to counteract biases in the processes by which articles come to exist (or are deleted~\cite{tripodi2021ms})---recommending articles that should exist but don't is an interesting area for future study.
        
        \item \textbf{Fairness constructs}: we focus on gender and geography in this challenge as two metrics for which there is high data coverage and clearer expectations about what "fairer" or more representative coverage might look like. That does not mean these are the most important constructs, but others---e.g., religion, sexuality, culture, race---generally are either more challenging to model or map to fairness goals~\cite{redi2020taxonomy}.
    \end{itemize}
\end{itemize}

%% file: nb-alignments.tex
\hypertarget{alignments}{%
\section{Alignments}\label{alignments}}

This appendix provides further details on how the page alignments and target distributions are computed.  It is a Jupyter notebook analyzes page alignments and prepares metrics for final
use. It needs to be run to create the serialized alignment data files
the metrics require; it is available in the code that goes with the appendix.

Its final output is \textbf{pickled metric objects}: an instance of the
Task 1 and Task 2 metric classes, serialized to a compressed file with
\href{https://binpickle.lenskit.org}{binpickle}.

\hypertarget{setup}{%
\subsection{Setup}\label{setup}}

We begin by loading necessary libraries:

\begin{Shaded}
\begin{Highlighting}[]
\ImportTok{from}\NormalTok{ pathlib }\ImportTok{import}\NormalTok{ Path}
\ImportTok{import}\NormalTok{ pandas }\ImportTok{as}\NormalTok{ pd}
\ImportTok{import}\NormalTok{ xarray }\ImportTok{as}\NormalTok{ xr}
\ImportTok{import}\NormalTok{ numpy }\ImportTok{as}\NormalTok{ np}
\ImportTok{import}\NormalTok{ matplotlib.pyplot }\ImportTok{as}\NormalTok{ plt}
\ImportTok{import}\NormalTok{ seaborn }\ImportTok{as}\NormalTok{ sns}
\ImportTok{import}\NormalTok{ gzip}
\ImportTok{import}\NormalTok{ pickle}
\ImportTok{import}\NormalTok{ binpickle}
\ImportTok{from}\NormalTok{ natural.size }\ImportTok{import}\NormalTok{ binarysize}
\end{Highlighting}
\end{Shaded}

We're going to use ZStandard compression to save our metrics, so let's
create a codec object:

\begin{Shaded}
\begin{Highlighting}[]
\NormalTok{codec }\OperatorTok{=}\NormalTok{ binpickle.codecs.Blosc(}\StringTok{\textquotesingle{}zstd\textquotesingle{}}\NormalTok{)}
\end{Highlighting}
\end{Shaded}

Set up progress bar and logging support:

\begin{Shaded}
\begin{Highlighting}[]
\ImportTok{from}\NormalTok{ tqdm.auto }\ImportTok{import}\NormalTok{ tqdm}
\NormalTok{tqdm.pandas(leave}\OperatorTok{=}\VariableTok{False}\NormalTok{)}
\end{Highlighting}
\end{Shaded}

\begin{Shaded}
\begin{Highlighting}[]
\ImportTok{import}\NormalTok{ sys, logging}
\NormalTok{logging.basicConfig(level}\OperatorTok{=}\NormalTok{logging.INFO, stream}\OperatorTok{=}\NormalTok{sys.stderr)}
\NormalTok{log }\OperatorTok{=}\NormalTok{ logging.getLogger(}\StringTok{\textquotesingle{}alignment\textquotesingle{}}\NormalTok{)}
\end{Highlighting}
\end{Shaded}

Import metric code:

\begin{Shaded}
\begin{Highlighting}[]
\OperatorTok{\%}\NormalTok{load\_ext autoreload}
\OperatorTok{\%}\NormalTok{autoreload }\DecValTok{1}
\end{Highlighting}
\end{Shaded}

\begin{Shaded}
\begin{Highlighting}[]
\OperatorTok{\%}\NormalTok{aimport metrics}
\ImportTok{from}\NormalTok{ trecdata }\ImportTok{import}\NormalTok{ scan\_runs}
\end{Highlighting}
\end{Shaded}

\hypertarget{loading-data}{%
\subsection{Loading Data}\label{loading-data}}

We first load the page metadata:

\begin{Shaded}
\begin{Highlighting}[]
\NormalTok{pages }\OperatorTok{=}\NormalTok{ pd.read\_json(}\StringTok{\textquotesingle{}data/trec\_metadata\_eval.json.gz\textquotesingle{}}\NormalTok{, lines}\OperatorTok{=}\VariableTok{True}\NormalTok{)}
\NormalTok{pages }\OperatorTok{=}\NormalTok{ pages.drop\_duplicates(}\StringTok{\textquotesingle{}page\_id\textquotesingle{}}\NormalTok{)}
\NormalTok{pages.info()}
\end{Highlighting}
\end{Shaded}

\begin{verbatim}
<class 'pandas.core.frame.DataFrame'>
Int64Index: 6023415 entries, 0 to 6023435
Data columns (total 5 columns):
 #   Column                Dtype  
---  ------                -----  
 0   page_id               int64  
 1   quality_score         float64
 2   quality_score_disc    object 
 3   geographic_locations  object 
 4   gender                object 
dtypes: float64(1), int64(1), object(3)
memory usage: 275.7+ MB
\end{verbatim}

Now we will load the evaluation topics:

\begin{Shaded}
\begin{Highlighting}[]
\NormalTok{eval\_topics }\OperatorTok{=}\NormalTok{ pd.read\_json(}\StringTok{\textquotesingle{}data/eval{-}topics{-}with{-}qrels.json.gz\textquotesingle{}}\NormalTok{, lines}\OperatorTok{=}\VariableTok{True}\NormalTok{)}
\NormalTok{eval\_topics.info()}
\end{Highlighting}
\end{Shaded}

\begin{verbatim}
<class 'pandas.core.frame.DataFrame'>
RangeIndex: 49 entries, 0 to 48
Data columns (total 5 columns):
 #   Column         Non-Null Count  Dtype 
---  ------         --------------  ----- 
 0   id             49 non-null     int64 
 1   title          49 non-null     object
 2   rel_docs       49 non-null     object
 3   assessed_docs  49 non-null     object
 4   max_tier       49 non-null     int64 
dtypes: int64(2), object(3)
memory usage: 2.0+ KB
\end{verbatim}

\begin{Shaded}
\begin{Highlighting}[]
\NormalTok{train\_topics }\OperatorTok{=}\NormalTok{ pd.read\_json(}\StringTok{\textquotesingle{}data/trec\_topics.json.gz\textquotesingle{}}\NormalTok{, lines}\OperatorTok{=}\VariableTok{True}\NormalTok{)}
\NormalTok{train\_topics.info()}
\end{Highlighting}
\end{Shaded}

\begin{verbatim}
<class 'pandas.core.frame.DataFrame'>
RangeIndex: 57 entries, 0 to 56
Data columns (total 6 columns):
 #   Column    Non-Null Count  Dtype 
---  ------    --------------  ----- 
 0   id        57 non-null     int64 
 1   title     57 non-null     object
 2   keywords  57 non-null     object
 3   scope     57 non-null     object
 4   homepage  57 non-null     object
 5   rel_docs  57 non-null     object
dtypes: int64(1), object(5)
memory usage: 2.8+ KB
\end{verbatim}

Train and eval topics use a disjoint set of IDs:

\begin{Shaded}
\begin{Highlighting}[]
\NormalTok{train\_topics[}\StringTok{\textquotesingle{}id\textquotesingle{}}\NormalTok{].describe()}
\end{Highlighting}
\end{Shaded}

\begin{verbatim}
count    57.000000
mean     29.000000
std      16.598193
min       1.000000
25%      15.000000
50%      29.000000
75%      43.000000
max      57.000000
Name: id, dtype: float64
\end{verbatim}

\begin{Shaded}
\begin{Highlighting}[]
\NormalTok{eval\_topics[}\StringTok{\textquotesingle{}id\textquotesingle{}}\NormalTok{].describe()}
\end{Highlighting}
\end{Shaded}

\begin{verbatim}
count     49.000000
mean     125.346939
std       14.687794
min      101.000000
25%      113.000000
50%      125.000000
75%      138.000000
max      150.000000
Name: id, dtype: float64
\end{verbatim}

This allows us to create a single, integrated topics list for
convenience:

\begin{Shaded}
\begin{Highlighting}[]
\NormalTok{topics }\OperatorTok{=}\NormalTok{ pd.concat([train\_topics, eval\_topics], ignore\_index}\OperatorTok{=}\VariableTok{True}\NormalTok{)}
\NormalTok{topics[}\StringTok{\textquotesingle{}eval\textquotesingle{}}\NormalTok{] }\OperatorTok{=} \VariableTok{False}
\NormalTok{topics.loc[topics[}\StringTok{\textquotesingle{}id\textquotesingle{}}\NormalTok{] }\OperatorTok{\textgreater{}=} \DecValTok{100}\NormalTok{, }\StringTok{\textquotesingle{}eval\textquotesingle{}}\NormalTok{] }\OperatorTok{=} \VariableTok{True}
\NormalTok{topics.head()}
\end{Highlighting}
\end{Shaded}

\begin{verbatim}
   id         title                                           keywords  \
0   1   Agriculture  [agriculture, crops, livestock, forests, farming]   
1   2  Architecture  [architecture, skyscraper, landscape, building...   
2   3     Athletics      [athletics, player, sports, game, gymnastics]   
3   4      Aviation  [aviation, aircraft, airplane, airship, pilot,...   
4   5      Baseball                                         [baseball]   

                                               scope  \
0  This WikiProject strives to develop and improv...   
1  This WikiProject aims to: 1. Thoroughly explor...   
2  WikiProject Athletics, a project focused on im...   
3  The project generally considers any article re...   
4  Articles pertaining to baseball including base...   

                                            homepage  \
0  https://en.wikipedia.org/wiki/Wikipedia:WikiPr...   
1  https://en.wikipedia.org/wiki/Wikipedia:WikiPr...   
2  https://en.wikipedia.org/wiki/Wikipedia:WikiPr...   
3  https://en.wikipedia.org/wiki/Wikipedia:WikiPr...   
4  https://en.wikipedia.org/wiki/Wikipedia:WikiPr...   

                                            rel_docs assessed_docs  max_tier  \
0  [572, 627, 903, 1193, 1542, 1634, 3751, 3866, ...           NaN       NaN   
1  [682, 954, 1170, 1315, 1322, 1324, 1325, 1435,...           NaN       NaN   
2  [5729, 8490, 9623, 10391, 12231, 13791, 16078,...           NaN       NaN   
3  [849, 852, 1293, 1902, 1942, 2039, 2075, 2082,...           NaN       NaN   
4  [1135, 1136, 1293, 1893, 2129, 2140, 3797, 380...           NaN       NaN   

    eval  
0  False  
1  False  
2  False  
3  False  
4  False  
\end{verbatim}

Finally, a bit of hard-coded data - the world population:

\begin{Shaded}
\begin{Highlighting}[]
\NormalTok{world\_pop }\OperatorTok{=}\NormalTok{ pd.Series(\{}
    \StringTok{\textquotesingle{}Africa\textquotesingle{}}\NormalTok{: }\FloatTok{0.155070563}\NormalTok{,}
    \StringTok{\textquotesingle{}Antarctica\textquotesingle{}}\NormalTok{: }\FloatTok{1.54424E{-}07}\NormalTok{,}
    \StringTok{\textquotesingle{}Asia\textquotesingle{}}\NormalTok{: }\FloatTok{0.600202585}\NormalTok{,}
    \StringTok{\textquotesingle{}Europe\textquotesingle{}}\NormalTok{: }\FloatTok{0.103663858}\NormalTok{,}
    \StringTok{\textquotesingle{}Latin America and the Caribbean\textquotesingle{}}\NormalTok{: }\FloatTok{0.08609797}\NormalTok{,}
    \StringTok{\textquotesingle{}Northern America\textquotesingle{}}\NormalTok{: }\FloatTok{0.049616733}\NormalTok{,}
    \StringTok{\textquotesingle{}Oceania\textquotesingle{}}\NormalTok{: }\FloatTok{0.005348137}\NormalTok{,}
\NormalTok{\})}
\NormalTok{world\_pop.name }\OperatorTok{=} \StringTok{\textquotesingle{}geography\textquotesingle{}}
\end{Highlighting}
\end{Shaded}

And a gender global target:

\begin{Shaded}
\begin{Highlighting}[]
\NormalTok{gender\_tgt }\OperatorTok{=}\NormalTok{ pd.Series(\{}
    \StringTok{\textquotesingle{}female\textquotesingle{}}\NormalTok{: }\FloatTok{0.495}\NormalTok{,}
    \StringTok{\textquotesingle{}male\textquotesingle{}}\NormalTok{: }\FloatTok{0.495}\NormalTok{,}
    \StringTok{\textquotesingle{}third\textquotesingle{}}\NormalTok{: }\FloatTok{0.01}
\NormalTok{\})}
\NormalTok{gender\_tgt.name }\OperatorTok{=} \StringTok{\textquotesingle{}gender\textquotesingle{}}
\NormalTok{gender\_tgt.}\BuiltInTok{sum}\NormalTok{()}
\end{Highlighting}
\end{Shaded}

\begin{verbatim}
1.0
\end{verbatim}

Xarray intesectional global target:

\begin{Shaded}
\begin{Highlighting}[]
\NormalTok{geo\_tgt\_xa }\OperatorTok{=}\NormalTok{ xr.DataArray(world\_pop, dims}\OperatorTok{=}\NormalTok{[}\StringTok{\textquotesingle{}geography\textquotesingle{}}\NormalTok{])}
\NormalTok{gender\_tgt\_xa }\OperatorTok{=}\NormalTok{ xr.DataArray(gender\_tgt, dims}\OperatorTok{=}\NormalTok{[}\StringTok{\textquotesingle{}gender\textquotesingle{}}\NormalTok{])}
\NormalTok{int\_tgt }\OperatorTok{=}\NormalTok{ geo\_tgt\_xa }\OperatorTok{*}\NormalTok{ gender\_tgt\_xa}
\NormalTok{int\_tgt}
\end{Highlighting}
\end{Shaded}

\begin{verbatim}
<xarray.DataArray (geography: 7, gender: 3)>
array([[7.67599287e-02, 7.67599287e-02, 1.55070563e-03],
       [7.64398800e-08, 7.64398800e-08, 1.54424000e-09],
       [2.97100280e-01, 2.97100280e-01, 6.00202585e-03],
       [5.13136097e-02, 5.13136097e-02, 1.03663858e-03],
       [4.26184951e-02, 4.26184951e-02, 8.60979700e-04],
       [2.45602828e-02, 2.45602828e-02, 4.96167330e-04],
       [2.64732781e-03, 2.64732781e-03, 5.34813700e-05]])
Coordinates:
  * geography  (geography) object 'Africa' 'Antarctica' ... 'Oceania'
  * gender     (gender) object 'female' 'male' 'third'
\end{verbatim}

And the order of work-needed codes:

\begin{Shaded}
\begin{Highlighting}[]
\NormalTok{work\_order }\OperatorTok{=}\NormalTok{ [}
    \StringTok{\textquotesingle{}Stub\textquotesingle{}}\NormalTok{,}
    \StringTok{\textquotesingle{}Start\textquotesingle{}}\NormalTok{,}
    \StringTok{\textquotesingle{}C\textquotesingle{}}\NormalTok{,}
    \StringTok{\textquotesingle{}B\textquotesingle{}}\NormalTok{,}
    \StringTok{\textquotesingle{}GA\textquotesingle{}}\NormalTok{,}
    \StringTok{\textquotesingle{}FA\textquotesingle{}}\NormalTok{,}
\NormalTok{]}
\end{Highlighting}
\end{Shaded}

Now all our background data is set up.

\hypertarget{query-relevance}{%
\subsection{Query Relevance}\label{query-relevance}}

We now need to get the qrels for the topics. This is done by creating
frames with entries for every relevant document; missing documents are
assumed irrelevant (0).

In the individual metric evaluation files, we will truncate each run to
only the assessed documents (with a small amount of noise), so this is a
safe way to compute.

First the training topics:

\begin{Shaded}
\begin{Highlighting}[]
\NormalTok{train\_qrels }\OperatorTok{=}\NormalTok{ train\_topics[[}\StringTok{\textquotesingle{}id\textquotesingle{}}\NormalTok{, }\StringTok{\textquotesingle{}rel\_docs\textquotesingle{}}\NormalTok{]].explode(}\StringTok{\textquotesingle{}rel\_docs\textquotesingle{}}\NormalTok{, ignore\_index}\OperatorTok{=}\VariableTok{True}\NormalTok{)}
\NormalTok{train\_qrels.rename(columns}\OperatorTok{=}\NormalTok{\{}\StringTok{\textquotesingle{}rel\_docs\textquotesingle{}}\NormalTok{: }\StringTok{\textquotesingle{}page\_id\textquotesingle{}}\NormalTok{\}, inplace}\OperatorTok{=}\VariableTok{True}\NormalTok{)}
\NormalTok{train\_qrels[}\StringTok{\textquotesingle{}page\_id\textquotesingle{}}\NormalTok{] }\OperatorTok{=}\NormalTok{ train\_qrels[}\StringTok{\textquotesingle{}page\_id\textquotesingle{}}\NormalTok{].astype(}\StringTok{\textquotesingle{}i4\textquotesingle{}}\NormalTok{)}
\NormalTok{train\_qrels }\OperatorTok{=}\NormalTok{ train\_qrels.drop\_duplicates()}
\NormalTok{train\_qrels.head()}
\end{Highlighting}
\end{Shaded}

\begin{verbatim}
   id  page_id
0   1      572
1   1      627
2   1      903
3   1     1193
4   1     1542
\end{verbatim}

\begin{Shaded}
\begin{Highlighting}[]
\NormalTok{eval\_qrels }\OperatorTok{=}\NormalTok{ eval\_topics[[}\StringTok{\textquotesingle{}id\textquotesingle{}}\NormalTok{, }\StringTok{\textquotesingle{}rel\_docs\textquotesingle{}}\NormalTok{]].explode(}\StringTok{\textquotesingle{}rel\_docs\textquotesingle{}}\NormalTok{, ignore\_index}\OperatorTok{=}\VariableTok{True}\NormalTok{)}
\NormalTok{eval\_qrels.rename(columns}\OperatorTok{=}\NormalTok{\{}\StringTok{\textquotesingle{}rel\_docs\textquotesingle{}}\NormalTok{: }\StringTok{\textquotesingle{}page\_id\textquotesingle{}}\NormalTok{\}, inplace}\OperatorTok{=}\VariableTok{True}\NormalTok{)}
\NormalTok{eval\_qrels[}\StringTok{\textquotesingle{}page\_id\textquotesingle{}}\NormalTok{] }\OperatorTok{=}\NormalTok{ eval\_qrels[}\StringTok{\textquotesingle{}page\_id\textquotesingle{}}\NormalTok{].astype(}\StringTok{\textquotesingle{}i4\textquotesingle{}}\NormalTok{)}
\NormalTok{eval\_qrels }\OperatorTok{=}\NormalTok{ eval\_qrels.drop\_duplicates()}
\NormalTok{eval\_qrels.head()}
\end{Highlighting}
\end{Shaded}

\begin{verbatim}
    id  page_id
0  101      915
1  101     2948
2  101     9110
3  101     9742
4  101    10996
\end{verbatim}

And concatenate:

\begin{Shaded}
\begin{Highlighting}[]
\NormalTok{qrels }\OperatorTok{=}\NormalTok{ pd.concat([train\_qrels, eval\_qrels], ignore\_index}\OperatorTok{=}\VariableTok{True}\NormalTok{)}
\end{Highlighting}
\end{Shaded}

\hypertarget{page-alignments}{%
\subsection{Page Alignments}\label{page-alignments}}

All of our metrics require page "alignments": the protected-group
membership of each page.

\hypertarget{geography}{%
\subsubsection{Geography}\label{geography}}

Let's start with the straight page geography alignment for the public
evaluation of the training queries. The page metadata has that; let's
get the geography column.

\begin{Shaded}
\begin{Highlighting}[]
\NormalTok{page\_geo }\OperatorTok{=}\NormalTok{ pages[[}\StringTok{\textquotesingle{}page\_id\textquotesingle{}}\NormalTok{, }\StringTok{\textquotesingle{}geographic\_locations\textquotesingle{}}\NormalTok{]].explode(}\StringTok{\textquotesingle{}geographic\_locations\textquotesingle{}}\NormalTok{, ignore\_index}\OperatorTok{=}\VariableTok{True}\NormalTok{)}
\NormalTok{page\_geo.head()}
\end{Highlighting}
\end{Shaded}

\begin{verbatim}
   page_id geographic_locations
0       12                  NaN
1       25                  NaN
2       39                  NaN
3      290                  NaN
4      303     Northern America
\end{verbatim}

And we will now pivot this into a matrix so we get page alignment
vectors:

\begin{Shaded}
\begin{Highlighting}[]
\NormalTok{page\_geo\_align }\OperatorTok{=}\NormalTok{ page\_geo.assign(x}\OperatorTok{=}\DecValTok{1}\NormalTok{).pivot(index}\OperatorTok{=}\StringTok{\textquotesingle{}page\_id\textquotesingle{}}\NormalTok{, columns}\OperatorTok{=}\StringTok{\textquotesingle{}geographic\_locations\textquotesingle{}}\NormalTok{, values}\OperatorTok{=}\StringTok{\textquotesingle{}x\textquotesingle{}}\NormalTok{)}
\NormalTok{page\_geo\_align.rename(columns}\OperatorTok{=}\NormalTok{\{np.nan: }\StringTok{\textquotesingle{}Unknown\textquotesingle{}}\NormalTok{\}, inplace}\OperatorTok{=}\VariableTok{True}\NormalTok{)}
\NormalTok{page\_geo\_align.fillna(}\DecValTok{0}\NormalTok{, inplace}\OperatorTok{=}\VariableTok{True}\NormalTok{)}
\NormalTok{page\_geo\_align.head()}
\end{Highlighting}
\end{Shaded}

\begin{verbatim}
geographic_locations  Unknown  Africa  Antarctica  Asia  Europe  \
page_id                                                           
12                        1.0     0.0         0.0   0.0     0.0   
25                        1.0     0.0         0.0   0.0     0.0   
39                        1.0     0.0         0.0   0.0     0.0   
290                       1.0     0.0         0.0   0.0     0.0   
303                       0.0     0.0         0.0   0.0     0.0   

geographic_locations  Latin America and the Caribbean  Northern America  \
page_id                                                                   
12                                                0.0               0.0   
25                                                0.0               0.0   
39                                                0.0               0.0   
290                                               0.0               0.0   
303                                               0.0               1.0   

geographic_locations  Oceania  
page_id                        
12                        0.0  
25                        0.0  
39                        0.0  
290                       0.0  
303                       0.0  
\end{verbatim}

And convert this to an xarray for multidimensional usage:

\begin{Shaded}
\begin{Highlighting}[]
\NormalTok{page\_geo\_xr }\OperatorTok{=}\NormalTok{ xr.DataArray(page\_geo\_align, dims}\OperatorTok{=}\NormalTok{[}\StringTok{\textquotesingle{}page\textquotesingle{}}\NormalTok{, }\StringTok{\textquotesingle{}geography\textquotesingle{}}\NormalTok{])}
\NormalTok{page\_geo\_xr}
\end{Highlighting}
\end{Shaded}

\begin{verbatim}
<xarray.DataArray (page: 6023415, geography: 8)>
array([[1., 0., 0., ..., 0., 0., 0.],
       [1., 0., 0., ..., 0., 0., 0.],
       [1., 0., 0., ..., 0., 0., 0.],
       ...,
       [1., 0., 0., ..., 0., 0., 0.],
       [1., 0., 0., ..., 0., 0., 0.],
       [1., 0., 0., ..., 0., 0., 0.]])
Coordinates:
  * page       (page) int64 12 25 39 290 ... 67268663 67268668 67268699 67268751
  * geography  (geography) object 'Unknown' 'Africa' ... 'Oceania'
\end{verbatim}

\begin{Shaded}
\begin{Highlighting}[]
\NormalTok{binarysize(page\_geo\_xr.nbytes)}
\end{Highlighting}
\end{Shaded}

\begin{verbatim}
'385.50 MiB'
\end{verbatim}

\hypertarget{gender}{%
\subsubsection{Gender}\label{gender}}

The "undisclosed personal attribute" is gender. Not all articles have
gender as a relevant variable - articles not about a living being
generally will not.

We're going to follow the same approach for gender:

\begin{Shaded}
\begin{Highlighting}[]
\NormalTok{page\_gender }\OperatorTok{=}\NormalTok{ pages[[}\StringTok{\textquotesingle{}page\_id\textquotesingle{}}\NormalTok{, }\StringTok{\textquotesingle{}gender\textquotesingle{}}\NormalTok{]].explode(}\StringTok{\textquotesingle{}gender\textquotesingle{}}\NormalTok{, ignore\_index}\OperatorTok{=}\VariableTok{True}\NormalTok{)}
\NormalTok{page\_gender.fillna(}\StringTok{\textquotesingle{}unknown\textquotesingle{}}\NormalTok{, inplace}\OperatorTok{=}\VariableTok{True}\NormalTok{)}
\NormalTok{page\_gender.head()}
\end{Highlighting}
\end{Shaded}

\begin{verbatim}
   page_id   gender
0       12  unknown
1       25  unknown
2       39  unknown
3      290  unknown
4      303  unknown
\end{verbatim}

We need to do a little targeted repair - there is an erroneous record of
a gender of "Taira no Kiyomori" is actually male. Replace that:

\begin{Shaded}
\begin{Highlighting}[]
\NormalTok{page\_gender }\OperatorTok{=}\NormalTok{ page\_gender.loc[page\_gender[}\StringTok{\textquotesingle{}gender\textquotesingle{}}\NormalTok{] }\OperatorTok{!=} \StringTok{\textquotesingle{}Taira no Kiyomori\textquotesingle{}}\NormalTok{]}
\end{Highlighting}
\end{Shaded}

Now, we're going to do a little more work to reduce the dimensionality
of the space. Points:

\begin{enumerate}
\tightlist
\item
  Trans men are men
\item
  Trans women are women
\item
  Cisgender is an adjective that can be dropped for the present purposes
\end{enumerate}

The result is that we will collapse "transgender female" and "cisgender
female" into "female".

The \textbf{downside} to this is that trans men are probabily
significantly under-represented, but are now being collapsed into the
dominant group.

\begin{Shaded}
\begin{Highlighting}[]
\NormalTok{pgcol }\OperatorTok{=}\NormalTok{ page\_gender[}\StringTok{\textquotesingle{}gender\textquotesingle{}}\NormalTok{]}
\NormalTok{pgcol }\OperatorTok{=}\NormalTok{ pgcol.}\BuiltInTok{str}\NormalTok{.replace(}\VerbatimStringTok{r\textquotesingle{}(?:tran|ci)sgender\textbackslash{}s+((?:fe)?male)\textquotesingle{}}\NormalTok{, }\VerbatimStringTok{r\textquotesingle{}\textbackslash{}1\textquotesingle{}}\NormalTok{, regex}\OperatorTok{=}\VariableTok{True}\NormalTok{)}
\end{Highlighting}
\end{Shaded}

Now, we're going to group the remaining gender identities together under
the label 'third'. As noted above, this is a debatable exercise that
collapses a lot of identity.

\begin{Shaded}
\begin{Highlighting}[]
\NormalTok{genders }\OperatorTok{=}\NormalTok{ [}\StringTok{\textquotesingle{}unknown\textquotesingle{}}\NormalTok{, }\StringTok{\textquotesingle{}male\textquotesingle{}}\NormalTok{, }\StringTok{\textquotesingle{}female\textquotesingle{}}\NormalTok{, }\StringTok{\textquotesingle{}third\textquotesingle{}}\NormalTok{]}
\NormalTok{pgcol[}\OperatorTok{\textasciitilde{}}\NormalTok{pgcol.isin(genders)] }\OperatorTok{=} \StringTok{\textquotesingle{}third\textquotesingle{}}
\end{Highlighting}
\end{Shaded}

Now put this column back in the frame and deduplicate.

\begin{Shaded}
\begin{Highlighting}[]
\NormalTok{page\_gender[}\StringTok{\textquotesingle{}gender\textquotesingle{}}\NormalTok{] }\OperatorTok{=}\NormalTok{ pgcol}
\NormalTok{page\_gender }\OperatorTok{=}\NormalTok{ page\_gender.drop\_duplicates()}
\end{Highlighting}
\end{Shaded}

And make an alignment matrix (reordering so 'unknown' is first for
consistency):

\begin{Shaded}
\begin{Highlighting}[]
\NormalTok{page\_gend\_align }\OperatorTok{=}\NormalTok{ page\_gender.assign(x}\OperatorTok{=}\DecValTok{1}\NormalTok{).pivot(index}\OperatorTok{=}\StringTok{\textquotesingle{}page\_id\textquotesingle{}}\NormalTok{, columns}\OperatorTok{=}\StringTok{\textquotesingle{}gender\textquotesingle{}}\NormalTok{, values}\OperatorTok{=}\StringTok{\textquotesingle{}x\textquotesingle{}}\NormalTok{)}
\NormalTok{page\_gend\_align.fillna(}\DecValTok{0}\NormalTok{, inplace}\OperatorTok{=}\VariableTok{True}\NormalTok{)}
\NormalTok{page\_gend\_align }\OperatorTok{=}\NormalTok{ page\_gend\_align.reindex(columns}\OperatorTok{=}\NormalTok{[}\StringTok{\textquotesingle{}unknown\textquotesingle{}}\NormalTok{, }\StringTok{\textquotesingle{}female\textquotesingle{}}\NormalTok{, }\StringTok{\textquotesingle{}male\textquotesingle{}}\NormalTok{, }\StringTok{\textquotesingle{}third\textquotesingle{}}\NormalTok{])}
\NormalTok{page\_gend\_align.head()}
\end{Highlighting}
\end{Shaded}

\begin{verbatim}
gender   unknown  female  male  third
page_id                              
12           1.0     0.0   0.0    0.0
25           1.0     0.0   0.0    0.0
39           1.0     0.0   0.0    0.0
290          1.0     0.0   0.0    0.0
303          1.0     0.0   0.0    0.0
\end{verbatim}

Let's see how frequent each of the genders is:

\begin{Shaded}
\begin{Highlighting}[]
\NormalTok{page\_gend\_align.}\BuiltInTok{sum}\NormalTok{(axis}\OperatorTok{=}\DecValTok{0}\NormalTok{).sort\_values(ascending}\OperatorTok{=}\VariableTok{False}\NormalTok{)}
\end{Highlighting}
\end{Shaded}

\begin{verbatim}
gender
unknown    4246540.0
male       1441813.0
female      334946.0
third          452.0
dtype: float64
\end{verbatim}

And convert to an xarray:

\begin{Shaded}
\begin{Highlighting}[]
\NormalTok{page\_gend\_xr }\OperatorTok{=}\NormalTok{ xr.DataArray(page\_gend\_align, dims}\OperatorTok{=}\NormalTok{[}\StringTok{\textquotesingle{}page\textquotesingle{}}\NormalTok{, }\StringTok{\textquotesingle{}gender\textquotesingle{}}\NormalTok{])}
\NormalTok{page\_gend\_xr}
\end{Highlighting}
\end{Shaded}

\begin{verbatim}
<xarray.DataArray (page: 6023415, gender: 4)>
array([[1., 0., 0., 0.],
       [1., 0., 0., 0.],
       [1., 0., 0., 0.],
       ...,
       [0., 1., 0., 0.],
       [1., 0., 0., 0.],
       [1., 0., 0., 0.]])
Coordinates:
  * page     (page) int64 12 25 39 290 ... 67268663 67268668 67268699 67268751
  * gender   (gender) object 'unknown' 'female' 'male' 'third'
\end{verbatim}

\begin{Shaded}
\begin{Highlighting}[]
\NormalTok{binarysize(page\_gend\_xr.nbytes)}
\end{Highlighting}
\end{Shaded}

\begin{verbatim}
'192.75 MiB'
\end{verbatim}

\hypertarget{intersectional-alignment}{%
\subsubsection{Intersectional
Alignment}\label{intersectional-alignment}}

We'll now convert this data array to an \textbf{intersectional}
alignment array:

\begin{Shaded}
\begin{Highlighting}[]
\NormalTok{page\_xalign }\OperatorTok{=}\NormalTok{ page\_geo\_xr }\OperatorTok{*}\NormalTok{ page\_gend\_xr}
\NormalTok{page\_xalign}
\end{Highlighting}
\end{Shaded}

\begin{verbatim}
<xarray.DataArray (page: 6023415, geography: 8, gender: 4)>
array([[[1., 0., 0., 0.],
        [0., 0., 0., 0.],
        [0., 0., 0., 0.],
        ...,
        [0., 0., 0., 0.],
        [0., 0., 0., 0.],
        [0., 0., 0., 0.]],

       [[1., 0., 0., 0.],
        [0., 0., 0., 0.],
        [0., 0., 0., 0.],
        ...,
        [0., 0., 0., 0.],
        [0., 0., 0., 0.],
        [0., 0., 0., 0.]],

       [[1., 0., 0., 0.],
        [0., 0., 0., 0.],
        [0., 0., 0., 0.],
        ...,
...
        ...,
        [0., 0., 0., 0.],
        [0., 0., 0., 0.],
        [0., 0., 0., 0.]],

       [[1., 0., 0., 0.],
        [0., 0., 0., 0.],
        [0., 0., 0., 0.],
        ...,
        [0., 0., 0., 0.],
        [0., 0., 0., 0.],
        [0., 0., 0., 0.]],

       [[1., 0., 0., 0.],
        [0., 0., 0., 0.],
        [0., 0., 0., 0.],
        ...,
        [0., 0., 0., 0.],
        [0., 0., 0., 0.],
        [0., 0., 0., 0.]]])
Coordinates:
  * page       (page) int64 12 25 39 290 ... 67268663 67268668 67268699 67268751
  * geography  (geography) object 'Unknown' 'Africa' ... 'Oceania'
  * gender     (gender) object 'unknown' 'female' 'male' 'third'
\end{verbatim}

\begin{Shaded}
\begin{Highlighting}[]
\NormalTok{binarysize(page\_xalign.nbytes)}
\end{Highlighting}
\end{Shaded}

\begin{verbatim}
'1.54 GiB'
\end{verbatim}

Make sure that did the right thing and we have intersectional numbers:

\begin{Shaded}
\begin{Highlighting}[]
\NormalTok{page\_xalign.}\BuiltInTok{sum}\NormalTok{(axis}\OperatorTok{=}\DecValTok{0}\NormalTok{)}
\end{Highlighting}
\end{Shaded}

\begin{verbatim}
<xarray.DataArray (geography: 8, gender: 4)>
array([[2.06922e+06, 8.21940e+04, 4.05772e+05, 1.85000e+02],
       [7.76580e+04, 1.04830e+04, 4.34670e+04, 8.00000e+00],
       [9.62500e+03, 0.00000e+00, 1.00000e+00, 0.00000e+00],
       [4.27422e+05, 3.79980e+04, 1.35310e+05, 2.10000e+01],
       [7.65203e+05, 9.67970e+04, 4.27747e+05, 6.30000e+01],
       [1.01464e+05, 1.61660e+04, 6.77640e+04, 4.00000e+00],
       [7.21244e+05, 8.25430e+04, 3.30205e+05, 1.59000e+02],
       [9.26820e+04, 1.45240e+04, 5.07260e+04, 2.00000e+01]])
Coordinates:
  * geography  (geography) object 'Unknown' 'Africa' ... 'Oceania'
  * gender     (gender) object 'unknown' 'female' 'male' 'third'
\end{verbatim}

And make sure combination with targets work as expected:

\begin{Shaded}
\begin{Highlighting}[]
\NormalTok{(page\_xalign.}\BuiltInTok{sum}\NormalTok{(axis}\OperatorTok{=}\DecValTok{0}\NormalTok{) }\OperatorTok{+}\NormalTok{ int\_tgt) }\OperatorTok{*} \FloatTok{0.5}
\end{Highlighting}
\end{Shaded}

\begin{verbatim}
<xarray.DataArray (geography: 7, gender: 3)>
array([[5.24153838e+03, 2.17335384e+04, 4.00077535e+00],
       [3.82199400e-08, 5.00000038e-01, 7.72120000e-10],
       [1.89991486e+04, 6.76551486e+04, 1.05030010e+01],
       [4.83985257e+04, 2.13873526e+05, 3.15005183e+01],
       [8.08302131e+03, 3.38820213e+04, 2.00043049e+00],
       [4.12715123e+04, 1.65102512e+05, 7.95002481e+01],
       [7.26200132e+03, 2.53630013e+04, 1.00000267e+01]])
Coordinates:
  * geography  (geography) object 'Africa' 'Antarctica' ... 'Oceania'
  * gender     (gender) object 'female' 'male' 'third'
\end{verbatim}

\hypertarget{task-1-metric-preparation}{%
\subsection{Task 1 Metric Preparation}\label{task-1-metric-preparation}}

Now that we have our alignments and qrels, we are ready to prepare the
Task 1 metrics.

Task 1 ignores the "unknown" alignment category, so we're going to
create a \texttt{kga} frame (for \textbf{K}nown \textbf{G}eographic
\textbf{A}lignment), and corresponding frames for intersectional
alignment.

\begin{Shaded}
\begin{Highlighting}[]
\NormalTok{page\_kga }\OperatorTok{=}\NormalTok{ page\_geo\_align.iloc[:, }\DecValTok{1}\NormalTok{:]}
\NormalTok{page\_kga.head()}
\end{Highlighting}
\end{Shaded}

\begin{verbatim}
geographic_locations  Africa  Antarctica  Asia  Europe  \
page_id                                                  
12                       0.0         0.0   0.0     0.0   
25                       0.0         0.0   0.0     0.0   
39                       0.0         0.0   0.0     0.0   
290                      0.0         0.0   0.0     0.0   
303                      0.0         0.0   0.0     0.0   

geographic_locations  Latin America and the Caribbean  Northern America  \
page_id                                                                   
12                                                0.0               0.0   
25                                                0.0               0.0   
39                                                0.0               0.0   
290                                               0.0               0.0   
303                                               0.0               1.0   

geographic_locations  Oceania  
page_id                        
12                        0.0  
25                        0.0  
39                        0.0  
290                       0.0  
303                       0.0  
\end{verbatim}

Intersectional is a little harder to do, because things can be
\textbf{intersectionally unknown}: we may know gender but not geography,
or vice versa. To deal with these missing values for Task 1, we're going
to ignore \emph{totally unknown} values, but keep partially-known as a
category.

We also need to ravel our tensors into a matrix for compatibility with
the metric code. Since 'unknown' is the first value on each axis, we can
ravel, and then drop the first column.

\begin{Shaded}
\begin{Highlighting}[]
\NormalTok{xshp }\OperatorTok{=}\NormalTok{ page\_xalign.shape}
\NormalTok{xshp }\OperatorTok{=}\NormalTok{ (xshp[}\DecValTok{0}\NormalTok{], xshp[}\DecValTok{1}\NormalTok{] }\OperatorTok{*}\NormalTok{ xshp[}\DecValTok{2}\NormalTok{])}
\NormalTok{page\_xa\_df }\OperatorTok{=}\NormalTok{ pd.DataFrame(page\_xalign.values.reshape(xshp), index}\OperatorTok{=}\NormalTok{page\_xalign.indexes[}\StringTok{\textquotesingle{}page\textquotesingle{}}\NormalTok{])}
\NormalTok{page\_xa\_df.head()}
\end{Highlighting}
\end{Shaded}

\begin{verbatim}
       0    1    2    3    4    5    6    7    8    9   ...   22   23   24  \
page                                                    ...                  
12    1.0  0.0  0.0  0.0  0.0  0.0  0.0  0.0  0.0  0.0  ...  0.0  0.0  0.0   
25    1.0  0.0  0.0  0.0  0.0  0.0  0.0  0.0  0.0  0.0  ...  0.0  0.0  0.0   
39    1.0  0.0  0.0  0.0  0.0  0.0  0.0  0.0  0.0  0.0  ...  0.0  0.0  0.0   
290   1.0  0.0  0.0  0.0  0.0  0.0  0.0  0.0  0.0  0.0  ...  0.0  0.0  0.0   
303   0.0  0.0  0.0  0.0  0.0  0.0  0.0  0.0  0.0  0.0  ...  0.0  0.0  1.0   

       25   26   27   28   29   30   31  
page                                     
12    0.0  0.0  0.0  0.0  0.0  0.0  0.0  
25    0.0  0.0  0.0  0.0  0.0  0.0  0.0  
39    0.0  0.0  0.0  0.0  0.0  0.0  0.0  
290   0.0  0.0  0.0  0.0  0.0  0.0  0.0  
303   0.0  0.0  0.0  0.0  0.0  0.0  0.0  

[5 rows x 32 columns]
\end{verbatim}

And drop unknown, to get our page alignment vectors:

\begin{Shaded}
\begin{Highlighting}[]
\NormalTok{page\_kia }\OperatorTok{=}\NormalTok{ page\_xa\_df.iloc[:, }\DecValTok{1}\NormalTok{:]}
\end{Highlighting}
\end{Shaded}

\hypertarget{geographic-alignment}{%
\subsubsection{Geographic Alignment}\label{geographic-alignment}}

We'll start with the metric configuration for public training data,
considering only geographic alignment. We configure the metric to do
this for both the training and the eval queries.

\hypertarget{training-queries}{%
\paragraph{Training Queries}\label{training-queries}}

\begin{Shaded}
\begin{Highlighting}[]
\NormalTok{train\_qalign }\OperatorTok{=}\NormalTok{ train\_qrels.join(page\_kga, on}\OperatorTok{=}\StringTok{\textquotesingle{}page\_id\textquotesingle{}}\NormalTok{).drop(columns}\OperatorTok{=}\NormalTok{[}\StringTok{\textquotesingle{}page\_id\textquotesingle{}}\NormalTok{]).groupby(}\StringTok{\textquotesingle{}id\textquotesingle{}}\NormalTok{).}\BuiltInTok{sum}\NormalTok{()}
\NormalTok{tqa\_sums }\OperatorTok{=}\NormalTok{ train\_qalign.}\BuiltInTok{sum}\NormalTok{(axis}\OperatorTok{=}\DecValTok{1}\NormalTok{)}
\NormalTok{train\_qalign }\OperatorTok{=}\NormalTok{ train\_qalign.divide(tqa\_sums, axis}\OperatorTok{=}\DecValTok{0}\NormalTok{)}
\end{Highlighting}
\end{Shaded}

\begin{Shaded}
\begin{Highlighting}[]
\NormalTok{train\_qalign.head()}
\end{Highlighting}
\end{Shaded}

\begin{verbatim}
      Africa  Antarctica      Asia    Europe  Latin America and the Caribbean  \
id                                                                              
1   0.049495     0.00000  0.121886  0.356566                         0.031650   
2   0.013388     0.00000  0.112008  0.574026                         0.026105   
3   0.109664     0.00000  0.125529  0.456033                         0.100040   
4   0.062495     0.00025  0.116161  0.327272                         0.079514   
5   0.000835     0.00000  0.065433  0.010149                         0.064755   

    Northern America   Oceania  
id                              
1           0.261616  0.178788  
2           0.228715  0.045758  
3           0.158419  0.050316  
4           0.369277  0.045032  
5           0.850192  0.008636  
\end{verbatim}

\begin{Shaded}
\begin{Highlighting}[]
\NormalTok{train\_qtarget }\OperatorTok{=}\NormalTok{ (train\_qalign }\OperatorTok{+}\NormalTok{ world\_pop) }\OperatorTok{*} \FloatTok{0.5}
\NormalTok{train\_qtarget.head()}
\end{Highlighting}
\end{Shaded}

\begin{verbatim}
      Africa    Antarctica      Asia    Europe  \
id                                               
1   0.102283  7.721200e-08  0.361044  0.230115   
2   0.084229  7.721200e-08  0.356105  0.338845   
3   0.132367  7.721200e-08  0.362866  0.279848   
4   0.108783  1.250113e-04  0.358182  0.215468   
5   0.077953  7.721200e-08  0.332818  0.056906   

    Latin America and the Caribbean  Northern America   Oceania  
id                                                               
1                          0.058874          0.155616  0.092068  
2                          0.056101          0.139166  0.025553  
3                          0.093069          0.104018  0.027832  
4                          0.082806          0.209447  0.025190  
5                          0.075427          0.449904  0.006992  
\end{verbatim}

And we can prepare a metric and save it:

\begin{Shaded}
\begin{Highlighting}[]
\NormalTok{t1\_train\_metric }\OperatorTok{=}\NormalTok{ metrics.Task1Metric(train\_qrels.set\_index(}\StringTok{\textquotesingle{}id\textquotesingle{}}\NormalTok{), page\_kga, train\_qtarget)}
\NormalTok{binpickle.dump(t1\_train\_metric, }\StringTok{\textquotesingle{}task1{-}train{-}geo{-}metric.bpk\textquotesingle{}}\NormalTok{, codec}\OperatorTok{=}\NormalTok{codec)}
\end{Highlighting}
\end{Shaded}

\begin{verbatim}
INFO:binpickle.write:pickled 337312647 bytes with 5 buffers
\end{verbatim}

\hypertarget{eval-queries}{%
\paragraph{Eval Queries}\label{eval-queries}}

Do the same thing for the eval data for a geo-only eval metric:

\begin{Shaded}
\begin{Highlighting}[]
\NormalTok{eval\_qalign }\OperatorTok{=}\NormalTok{ eval\_qrels.join(page\_kga, on}\OperatorTok{=}\StringTok{\textquotesingle{}page\_id\textquotesingle{}}\NormalTok{).drop(columns}\OperatorTok{=}\NormalTok{[}\StringTok{\textquotesingle{}page\_id\textquotesingle{}}\NormalTok{]).groupby(}\StringTok{\textquotesingle{}id\textquotesingle{}}\NormalTok{).}\BuiltInTok{sum}\NormalTok{()}
\NormalTok{eqa\_sums }\OperatorTok{=}\NormalTok{ eval\_qalign.}\BuiltInTok{sum}\NormalTok{(axis}\OperatorTok{=}\DecValTok{1}\NormalTok{)}
\NormalTok{eval\_qalign }\OperatorTok{=}\NormalTok{ eval\_qalign.divide(eqa\_sums, axis}\OperatorTok{=}\DecValTok{0}\NormalTok{)}
\NormalTok{eval\_qtarget }\OperatorTok{=}\NormalTok{ (eval\_qalign }\OperatorTok{+}\NormalTok{ world\_pop) }\OperatorTok{*} \FloatTok{0.5}
\NormalTok{t1\_eval\_metric }\OperatorTok{=}\NormalTok{ metrics.Task1Metric(eval\_qrels.set\_index(}\StringTok{\textquotesingle{}id\textquotesingle{}}\NormalTok{), page\_kga, eval\_qtarget)}
\NormalTok{binpickle.dump(t1\_eval\_metric, }\StringTok{\textquotesingle{}task1{-}eval{-}geo{-}metric.bpk\textquotesingle{}}\NormalTok{, codec}\OperatorTok{=}\NormalTok{codec)}
\end{Highlighting}
\end{Shaded}

\begin{verbatim}
INFO:binpickle.write:pickled 337312643 bytes with 5 buffers
\end{verbatim}

\hypertarget{intersectional-alignment}{%
\subsubsection{Intersectional
Alignment}\label{intersectional-alignment}}

Now we need to apply similar logic, but for the intersectional
(geography * gender) alignment.

As noted as above, we need to carefully handle the unknown cases.

\hypertarget{demo}{%
\paragraph{Demo}\label{demo}}

To demonstrate how the logic works, let's first work it out in cells for
one query (1).

What are its documents?

\begin{Shaded}
\begin{Highlighting}[]
\NormalTok{qdf }\OperatorTok{=}\NormalTok{ qrels[qrels[}\StringTok{\textquotesingle{}id\textquotesingle{}}\NormalTok{] }\OperatorTok{==} \DecValTok{1}\NormalTok{]}
\NormalTok{qdf.name }\OperatorTok{=} \DecValTok{1}
\NormalTok{qdf}
\end{Highlighting}
\end{Shaded}

\begin{verbatim}
      id   page_id
0      1       572
1      1       627
2      1       903
3      1      1193
4      1      1542
...   ..       ...
6959   1  67066971
6960   1  67075177
6961   1  67178925
6962   1  67190032
6963   1  67244439

[6964 rows x 2 columns]
\end{verbatim}

We can use these page IDs to get its alignments:

\begin{Shaded}
\begin{Highlighting}[]
\NormalTok{q\_xa }\OperatorTok{=}\NormalTok{ page\_xalign.loc[qdf[}\StringTok{\textquotesingle{}page\_id\textquotesingle{}}\NormalTok{].values, :, :]}
\NormalTok{q\_xa}
\end{Highlighting}
\end{Shaded}

\begin{verbatim}
<xarray.DataArray (page: 6964, geography: 8, gender: 4)>
array([[[1., 0., 0., 0.],
        [0., 0., 0., 0.],
        [0., 0., 0., 0.],
        ...,
        [0., 0., 0., 0.],
        [0., 0., 0., 0.],
        [0., 0., 0., 0.]],

       [[1., 0., 0., 0.],
        [0., 0., 0., 0.],
        [0., 0., 0., 0.],
        ...,
        [0., 0., 0., 0.],
        [0., 0., 0., 0.],
        [0., 0., 0., 0.]],

       [[1., 0., 0., 0.],
        [0., 0., 0., 0.],
        [0., 0., 0., 0.],
        ...,
...
        ...,
        [0., 0., 0., 0.],
        [0., 0., 0., 0.],
        [0., 0., 0., 0.]],

       [[1., 0., 0., 0.],
        [0., 0., 0., 0.],
        [0., 0., 0., 0.],
        ...,
        [0., 0., 0., 0.],
        [0., 0., 0., 0.],
        [0., 0., 0., 0.]],

       [[1., 0., 0., 0.],
        [0., 0., 0., 0.],
        [0., 0., 0., 0.],
        ...,
        [0., 0., 0., 0.],
        [0., 0., 0., 0.],
        [0., 0., 0., 0.]]])
Coordinates:
  * page       (page) int64 572 627 903 1193 ... 67178925 67190032 67244439
  * geography  (geography) object 'Unknown' 'Africa' ... 'Oceania'
  * gender     (gender) object 'unknown' 'female' 'male' 'third'
\end{verbatim}

Summing over the first axis ('page') will produce an alignment matrix:

\begin{Shaded}
\begin{Highlighting}[]
\NormalTok{q\_am }\OperatorTok{=}\NormalTok{ q\_xa.}\BuiltInTok{sum}\NormalTok{(axis}\OperatorTok{=}\DecValTok{0}\NormalTok{)}
\NormalTok{q\_am}
\end{Highlighting}
\end{Shaded}

\begin{verbatim}
<xarray.DataArray (geography: 8, gender: 4)>
array([[3767.,   52.,  200.,    0.],
       [ 128.,   12.,    7.,    0.],
       [   0.,    0.,    0.,    0.],
       [ 322.,   11.,   29.,    0.],
       [ 940.,   23.,   96.,    0.],
       [  79.,    8.,    7.,    0.],
       [ 618.,   28.,  131.,    0.],
       [ 484.,    6.,   41.,    0.]])
Coordinates:
  * geography  (geography) object 'Unknown' 'Africa' ... 'Oceania'
  * gender     (gender) object 'unknown' 'female' 'male' 'third'
\end{verbatim}

Now we need to do reset the (0,0) coordinate (full unknown), and
normalize to a proportion.

\begin{Shaded}
\begin{Highlighting}[]
\NormalTok{q\_am[}\DecValTok{0}\NormalTok{, }\DecValTok{0}\NormalTok{] }\OperatorTok{=} \DecValTok{0}
\NormalTok{q\_am }\OperatorTok{=}\NormalTok{ q\_am }\OperatorTok{/}\NormalTok{ q\_am.}\BuiltInTok{sum}\NormalTok{()}
\NormalTok{q\_am}
\end{Highlighting}
\end{Shaded}

\begin{verbatim}
<xarray.DataArray (geography: 8, gender: 4)>
array([[0.        , 0.01613904, 0.06207325, 0.        ],
       [0.03972688, 0.00372439, 0.00217256, 0.        ],
       [0.        , 0.        , 0.        , 0.        ],
       [0.09993793, 0.00341403, 0.00900062, 0.        ],
       [0.29174426, 0.00713842, 0.02979516, 0.        ],
       [0.02451893, 0.00248293, 0.00217256, 0.        ],
       [0.19180633, 0.00869025, 0.04065798, 0.        ],
       [0.15021726, 0.0018622 , 0.01272502, 0.        ]])
Coordinates:
  * geography  (geography) object 'Unknown' 'Africa' ... 'Oceania'
  * gender     (gender) object 'unknown' 'female' 'male' 'third'
\end{verbatim}

Ok, now we have to - very carefully - average with our target modifier.
There are three groups:

\begin{itemize}
\tightlist
\item
  known (use intersectional target)
\item
  known-geo (use geo target)
\item
  known-gender (use gender target)
\end{itemize}

For each of these, we need to respect the fraction of the total it
represents. Let's compute those fractions:

\begin{Shaded}
\begin{Highlighting}[]
\NormalTok{q\_fk\_all }\OperatorTok{=}\NormalTok{ q\_am[}\DecValTok{1}\NormalTok{:, }\DecValTok{1}\NormalTok{:].}\BuiltInTok{sum}\NormalTok{()}
\NormalTok{q\_fk\_geo }\OperatorTok{=}\NormalTok{ q\_am[}\DecValTok{1}\NormalTok{:, :}\DecValTok{1}\NormalTok{].}\BuiltInTok{sum}\NormalTok{()}
\NormalTok{q\_fk\_gen }\OperatorTok{=}\NormalTok{ q\_am[:}\DecValTok{1}\NormalTok{, }\DecValTok{1}\NormalTok{:].}\BuiltInTok{sum}\NormalTok{()}
\NormalTok{q\_fk\_all, q\_fk\_geo, q\_fk\_gen}
\end{Highlighting}
\end{Shaded}

\begin{verbatim}
(<xarray.DataArray ()>
 array(0.12383613),
 <xarray.DataArray ()>
 array(0.79795158),
 <xarray.DataArray ()>
 array(0.07821229))
\end{verbatim}

And now do some surgery. Weighted-average to incorporate the target for
fully-known:

\begin{Shaded}
\begin{Highlighting}[]
\NormalTok{q\_tm }\OperatorTok{=}\NormalTok{ q\_am.copy()}
\NormalTok{q\_tm[}\DecValTok{1}\NormalTok{:, }\DecValTok{1}\NormalTok{:] }\OperatorTok{*=} \FloatTok{0.5}
\NormalTok{q\_tm[}\DecValTok{1}\NormalTok{:, }\DecValTok{1}\NormalTok{:] }\OperatorTok{+=}\NormalTok{ int\_tgt }\OperatorTok{*} \FloatTok{0.5} \OperatorTok{*}\NormalTok{ q\_fk\_all}
\NormalTok{q\_tm}
\end{Highlighting}
\end{Shaded}

\begin{verbatim}
<xarray.DataArray (geography: 8, gender: 4)>
array([[0.00000000e+00, 1.61390441e-02, 6.20732464e-02, 0.00000000e+00],
       [3.97268777e-02, 6.61502352e-03, 5.83910794e-03, 9.60166894e-05],
       [0.00000000e+00, 4.73300933e-09, 4.73300933e-09, 9.56163501e-11],
       [9.99379268e-02, 2.01028882e-02, 2.28961843e-02, 3.71633817e-04],
       [2.91744258e-01, 6.74645100e-03, 1.80748185e-02, 6.41866532e-05],
       [2.45189323e-02, 3.88031961e-03, 3.72513649e-03, 5.33101956e-05],
       [1.91806331e-01, 5.86585240e-03, 2.18497134e-02, 3.07217202e-05],
       [1.50217256e-01, 1.09501611e-03, 6.52642517e-03, 3.31146285e-06]])
Coordinates:
  * geography  (geography) object 'Unknown' 'Africa' ... 'Oceania'
  * gender     (gender) object 'unknown' 'female' 'male' 'third'
\end{verbatim}

And for known-geo:

\begin{Shaded}
\begin{Highlighting}[]
\NormalTok{q\_tm[}\DecValTok{1}\NormalTok{:, :}\DecValTok{1}\NormalTok{] }\OperatorTok{*=} \FloatTok{0.5}
\NormalTok{q\_tm[}\DecValTok{1}\NormalTok{:, :}\DecValTok{1}\NormalTok{] }\OperatorTok{+=}\NormalTok{ geo\_tgt\_xa }\OperatorTok{*} \FloatTok{0.5} \OperatorTok{*}\NormalTok{ q\_fk\_geo}
\end{Highlighting}
\end{Shaded}

And known-gender:

\begin{Shaded}
\begin{Highlighting}[]
\NormalTok{q\_tm[:}\DecValTok{1}\NormalTok{, }\DecValTok{1}\NormalTok{:] }\OperatorTok{*=} \FloatTok{0.5}
\NormalTok{q\_tm[:}\DecValTok{1}\NormalTok{, }\DecValTok{1}\NormalTok{:] }\OperatorTok{+=}\NormalTok{ gender\_tgt\_xa }\OperatorTok{*} \FloatTok{0.5} \OperatorTok{*}\NormalTok{ q\_fk\_gen}
\end{Highlighting}
\end{Shaded}

\begin{Shaded}
\begin{Highlighting}[]
\NormalTok{q\_tm}
\end{Highlighting}
\end{Shaded}

\begin{verbatim}
<xarray.DataArray (geography: 8, gender: 4)>
array([[0.00000000e+00, 2.74270639e-02, 5.03941651e-02, 3.91061453e-04],
       [8.17328395e-02, 6.61502352e-03, 5.83910794e-03, 9.60166894e-05],
       [6.16114376e-08, 4.73300933e-09, 4.73300933e-09, 9.56163501e-11],
       [2.89435265e-01, 2.01028882e-02, 2.28961843e-02, 3.71633817e-04],
       [1.87231499e-01, 6.74645100e-03, 1.80748185e-02, 6.41866532e-05],
       [4.66104719e-02, 3.88031961e-03, 3.72513649e-03, 5.33101956e-05],
       [1.15699041e-01, 5.86585240e-03, 2.18497134e-02, 3.07217202e-05],
       [7.72424054e-02, 1.09501611e-03, 6.52642517e-03, 3.31146285e-06]])
Coordinates:
  * geography  (geography) object 'Unknown' 'Africa' ... 'Oceania'
  * gender     (gender) object 'unknown' 'female' 'male' 'third'
\end{verbatim}

Now we can unravel this and drop the first entry:

\begin{Shaded}
\begin{Highlighting}[]
\NormalTok{q\_tm.values.ravel()[}\DecValTok{1}\NormalTok{:]}
\end{Highlighting}
\end{Shaded}

\begin{verbatim}
array([2.74270639e-02, 5.03941651e-02, 3.91061453e-04, 8.17328395e-02,
       6.61502352e-03, 5.83910794e-03, 9.60166894e-05, 6.16114376e-08,
       4.73300933e-09, 4.73300933e-09, 9.56163501e-11, 2.89435265e-01,
       2.01028882e-02, 2.28961843e-02, 3.71633817e-04, 1.87231499e-01,
       6.74645100e-03, 1.80748185e-02, 6.41866532e-05, 4.66104719e-02,
       3.88031961e-03, 3.72513649e-03, 5.33101956e-05, 1.15699041e-01,
       5.86585240e-03, 2.18497134e-02, 3.07217202e-05, 7.72424054e-02,
       1.09501611e-03, 6.52642517e-03, 3.31146285e-06])
\end{verbatim}

\hypertarget{implementation}{%
\paragraph{Implementation}\label{implementation}}

Now, to do this for every query, we'll use a function that takes a data
frame for a query's relevant docs and performs all of the above
operations:

\begin{Shaded}
\begin{Highlighting}[]
\KeywordTok{def}\NormalTok{ query\_xalign(qdf):}
\NormalTok{    pages }\OperatorTok{=}\NormalTok{ qdf[}\StringTok{\textquotesingle{}page\_id\textquotesingle{}}\NormalTok{]}
\NormalTok{    pages }\OperatorTok{=}\NormalTok{ pages[pages.isin(page\_xalign.indexes[}\StringTok{\textquotesingle{}page\textquotesingle{}}\NormalTok{])]}
\NormalTok{    q\_xa }\OperatorTok{=}\NormalTok{ page\_xalign.loc[pages.values, :, :]}
\NormalTok{    q\_am }\OperatorTok{=}\NormalTok{ q\_xa.}\BuiltInTok{sum}\NormalTok{(axis}\OperatorTok{=}\DecValTok{0}\NormalTok{)}

    \CommentTok{\# clear and normalize}
\NormalTok{    q\_am[}\DecValTok{0}\NormalTok{, }\DecValTok{0}\NormalTok{] }\OperatorTok{=} \DecValTok{0}
\NormalTok{    q\_am }\OperatorTok{=}\NormalTok{ q\_am }\OperatorTok{/}\NormalTok{ q\_am.}\BuiltInTok{sum}\NormalTok{()}
    
    \CommentTok{\# compute fractions in each section}
\NormalTok{    q\_fk\_all }\OperatorTok{=}\NormalTok{ q\_am[}\DecValTok{1}\NormalTok{:, }\DecValTok{1}\NormalTok{:].}\BuiltInTok{sum}\NormalTok{()}
\NormalTok{    q\_fk\_geo }\OperatorTok{=}\NormalTok{ q\_am[}\DecValTok{1}\NormalTok{:, :}\DecValTok{1}\NormalTok{].}\BuiltInTok{sum}\NormalTok{()}
\NormalTok{    q\_fk\_gen }\OperatorTok{=}\NormalTok{ q\_am[:}\DecValTok{1}\NormalTok{, }\DecValTok{1}\NormalTok{:].}\BuiltInTok{sum}\NormalTok{()}
    
    \CommentTok{\# known average}
\NormalTok{    q\_am[}\DecValTok{1}\NormalTok{:, }\DecValTok{1}\NormalTok{:] }\OperatorTok{*=} \FloatTok{0.5}
\NormalTok{    q\_am[}\DecValTok{1}\NormalTok{:, }\DecValTok{1}\NormalTok{:] }\OperatorTok{+=}\NormalTok{ int\_tgt }\OperatorTok{*} \FloatTok{0.5} \OperatorTok{*}\NormalTok{ q\_fk\_all}
    
    \CommentTok{\# known{-}geo average}
\NormalTok{    q\_am[}\DecValTok{1}\NormalTok{:, :}\DecValTok{1}\NormalTok{] }\OperatorTok{*=} \FloatTok{0.5}
\NormalTok{    q\_am[}\DecValTok{1}\NormalTok{:, :}\DecValTok{1}\NormalTok{] }\OperatorTok{+=}\NormalTok{ geo\_tgt\_xa }\OperatorTok{*} \FloatTok{0.5} \OperatorTok{*}\NormalTok{ q\_fk\_geo}
    
    \CommentTok{\# known{-}gender average}
\NormalTok{    q\_am[:}\DecValTok{1}\NormalTok{, }\DecValTok{1}\NormalTok{:] }\OperatorTok{*=} \FloatTok{0.5}
\NormalTok{    q\_am[:}\DecValTok{1}\NormalTok{, }\DecValTok{1}\NormalTok{:] }\OperatorTok{+=}\NormalTok{ gender\_tgt\_xa }\OperatorTok{*} \FloatTok{0.5} \OperatorTok{*}\NormalTok{ q\_fk\_gen}
    
    \CommentTok{\# and return the result}
    \ControlFlowTok{return}\NormalTok{ pd.Series(q\_am.values.ravel()[}\DecValTok{1}\NormalTok{:])}
\end{Highlighting}
\end{Shaded}

\begin{Shaded}
\begin{Highlighting}[]
\NormalTok{query\_xalign(qdf)}
\end{Highlighting}
\end{Shaded}

\begin{verbatim}
0     2.742706e-02
1     5.039417e-02
2     3.910615e-04
3     8.173284e-02
4     6.615024e-03
5     5.839108e-03
6     9.601669e-05
7     6.161144e-08
8     4.733009e-09
9     4.733009e-09
10    9.561635e-11
11    2.894353e-01
12    2.010289e-02
13    2.289618e-02
14    3.716338e-04
15    1.872315e-01
16    6.746451e-03
17    1.807482e-02
18    6.418665e-05
19    4.661047e-02
20    3.880320e-03
21    3.725136e-03
22    5.331020e-05
23    1.156990e-01
24    5.865852e-03
25    2.184971e-02
26    3.072172e-05
27    7.724241e-02
28    1.095016e-03
29    6.526425e-03
30    3.311463e-06
dtype: float64
\end{verbatim}

Now with that function, we can compute the alignment vector for each
query.

\begin{Shaded}
\begin{Highlighting}[]
\NormalTok{train\_qtarget }\OperatorTok{=}\NormalTok{ train\_qrels.groupby(}\StringTok{\textquotesingle{}id\textquotesingle{}}\NormalTok{).}\BuiltInTok{apply}\NormalTok{(query\_xalign)}
\NormalTok{train\_qtarget}
\end{Highlighting}
\end{Shaded}

\begin{verbatim}
          0         1         2         3         4         5         6   \
id                                                                         
1   0.027427  0.050394  0.000391  0.081733  0.006615  0.005839  0.000096   
2   0.012235  0.032073  0.000232  0.073168  0.003571  0.003669  0.000070   
3   0.022553  0.035541  0.000292  0.023527  0.040981  0.059556  0.000574   
4   0.012472  0.029112  0.000209  0.094840  0.004409  0.004901  0.000086   
5   0.023416  0.063398  0.000436  0.020521  0.024932  0.025194  0.000504   
6   0.126820  0.201558  0.001691  0.000021  0.028097  0.030215  0.000519   
7   0.050837  0.115432  0.000836  0.000026  0.034747  0.051416  0.000646   
8   0.038785  0.054361  0.000521  0.000065  0.038044  0.038746  0.000702   
9   0.059002  0.157276  0.001087  0.005630  0.028051  0.056632  0.000554   
10  0.064617  0.137545  0.001016  0.046254  0.008038  0.008038  0.000162   
11  0.060435  0.128320  0.000979  0.029760  0.020073  0.021687  0.000358   
12  0.020151  0.038214  0.000332  0.007651  0.032456  0.032725  0.000653   
13  0.014332  0.035379  0.000250  0.009309  0.034996  0.038045  0.000655   
14  0.046850  0.041965  0.000561  0.041140  0.016396  0.016052  0.000299   
15  0.068267  0.151259  0.001103  0.000377  0.030622  0.032842  0.000601   
16  0.135784  0.156886  0.001904  0.009884  0.025627  0.025196  0.000448   
17  0.067934  0.174287  0.001217  0.013029  0.026528  0.051011  0.000503   
18  0.018251  0.036588  0.000276  0.030822  0.023188  0.027944  0.000445   
19  0.103573  0.097443  0.001247  0.013579  0.027784  0.025663  0.000485   
20  0.089270  0.071413  0.000807  0.038786  0.021002  0.020420  0.000334   
21  0.108101  0.211342  0.001605  0.027217  0.016228  0.017338  0.000296   
22  0.026161  0.059192  0.000429  0.003891  0.036267  0.043268  0.000671   
23  0.024937  0.066835  0.000461  0.027385  0.025748  0.048173  0.000503   
24  0.081419  0.177767  0.001498  0.068779  0.005003  0.005585  0.000101   
25  0.107975  0.121840  0.001155  0.008671  0.026618  0.026711  0.000511   
26  0.041788  0.043538  0.000835  0.068601  0.027419  0.033208  0.000323   
27  0.038810  0.078021  0.000587  0.055195  0.006552  0.006769  0.000132   
28  0.023529  0.051249  0.000376  0.000296  0.035349  0.035468  0.000714   
29  0.087081  0.191848  0.001402  0.031724  0.014463  0.016746  0.000289   
30  0.056876  0.090125  0.000739  0.018007  0.027345  0.030689  0.000509   
31  0.053926  0.104654  0.000819  0.058785  0.010576  0.010732  0.000158   
32  0.061350  0.076521  0.003383  0.000257  0.036019  0.035636  0.001173   
33  0.030389  0.053928  0.000424  0.061496  0.010157  0.009832  0.000194   
34  0.096244  0.136311  0.001169  0.022726  0.020403  0.022574  0.000389   
35  0.087547  0.111057  0.000998  0.017422  0.028479  0.026493  0.000460   
36  0.112926  0.177999  0.001500  0.033846  0.017653  0.017997  0.000286   
37  0.186952  0.406140  0.003590  0.019620  0.006586  0.008204  0.000132   
38  0.027033  0.049546  0.000385  0.018166  0.026696  0.030492  0.000534   
39  0.050328  0.062821  0.000569  0.013145  0.027890  0.028400  0.000558   
40  0.165541  0.274278  0.002251  0.035258  0.004510  0.005636  0.000086   
41  0.043914  0.084869  0.000647  0.009875  0.035748  0.047432  0.000617   
42  0.042045  0.077639  0.000601  0.072132  0.011772  0.012999  0.000206   
43  0.088816  0.236394  0.001634  0.045119  0.006863  0.007995  0.000139   
44  0.073139  0.110720  0.000924  0.008159  0.029961  0.033199  0.000677   
45  0.102279  0.175641  0.001615  0.010408  0.025755  0.026896  0.000464   
46  0.049735  0.096882  0.000772  0.057043  0.016799  0.021288  0.000285   
47  0.002770  0.006868  0.000048  0.065464  0.006435  0.006435  0.000126   
48  0.048879  0.130381  0.000901  0.000374  0.031816  0.055275  0.000632   
49  0.023329  0.043920  0.000338  0.008751  0.032725  0.037509  0.000638   
50  0.021747  0.035992  0.000290  0.006747  0.053491  0.085334  0.000706   
51  0.042260  0.067426  0.000551  0.013784  0.027519  0.028445  0.000553   
52  0.011937  0.020075  0.000161  0.067402  0.003922  0.003828  0.000077   
53  0.064740  0.088902  0.000772  0.019943  0.029669  0.032655  0.000507   
54  0.006491  0.008134  0.000073  0.071422  0.004520  0.004449  0.000084   
55  0.001431  0.003218  0.000023  0.106429  0.000485  0.000485  0.000010   
56  0.127312  0.042703  0.000891  0.003167  0.049226  0.030645  0.000626   
57  0.064556  0.101994  0.000905  0.243893  0.072830  0.147165  0.000380   

              7             8             9   ...        21        22  \
id                                            ...                       
1   6.161144e-08  4.733009e-09  4.733009e-09  ...  0.003725  0.000053   
2   6.684003e-08  3.431828e-09  3.431828e-09  ...  0.003276  0.000039   
3   1.665751e-08  2.774295e-08  2.774295e-08  ...  0.039630  0.000312   
4   1.197781e-04  4.231748e-09  4.231748e-09  ...  0.002998  0.000048   
5   2.031701e-08  2.482831e-08  2.482831e-08  ...  0.038382  0.000280   
6   2.098131e-11  2.559433e-08  5.057750e-06  ...  0.019846  0.000288   
7   2.510823e-11  3.182076e-08  3.182076e-08  ...  0.052265  0.000358   
8   6.442770e-11  3.460808e-08  3.460808e-08  ...  0.029176  0.000396   
9   5.304181e-09  2.728670e-08  2.728670e-08  ...  0.074846  0.000307   
10  4.535438e-08  8.004062e-09  8.004062e-09  ...  0.005030  0.000090   
11  2.692777e-08  1.763909e-08  1.763909e-08  ...  0.015362  0.000199   
12  7.618901e-09  3.220520e-08  3.220520e-08  ...  0.018456  0.000363   
13  8.094806e-09  3.230352e-08  3.230352e-08  ...  0.042302  0.000364   
14  4.055074e-08  1.473139e-08  1.473139e-08  ...  0.009435  0.000166   
15  2.900724e-10  2.964395e-08  2.964395e-08  ...  0.031470  0.000334   
16  9.842701e-09  2.208922e-08  2.208922e-08  ...  0.015761  0.000249   
17  8.276972e-09  2.481865e-08  2.481865e-08  ...  0.024976  0.000280   
18  2.861625e-08  2.194840e-08  2.194840e-08  ...  0.034100  0.000247   
19  1.328731e-08  2.391224e-08  2.391224e-08  ...  0.018990  0.000269   
20  3.147228e-08  1.646900e-08  1.646900e-08  ...  0.009959  0.000185   
21  2.290303e-08  1.461251e-08  1.461251e-08  ...  0.012144  0.000165   
22  3.776095e-09  3.307219e-08  3.307219e-08  ...  0.056006  0.000373   
23  1.946374e-06  2.469539e-08  2.469539e-08  ...  0.062443  0.000278   
24  9.717212e-05  4.981659e-09  4.981659e-09  ...  0.004040  0.000056   
25  8.449778e-09  2.520965e-08  2.520965e-08  ...  0.019174  0.000284   
26  3.841889e-08  1.590954e-08  1.590954e-08  ...  0.018563  0.000179   
27  5.496504e-08  6.524548e-09  6.524548e-09  ...  0.040466  0.000073   
28  2.952129e-10  3.520143e-08  3.520143e-08  ...  0.020403  0.000396   
29  2.675860e-08  1.426019e-08  1.426019e-08  ...  0.010663  0.000161   
30  1.508983e-08  2.510388e-08  2.510388e-08  ...  0.023381  0.000283   
31  4.918211e-08  7.782549e-09  7.782549e-09  ...  0.006588  0.000088   
32  2.559296e-10  3.269452e-08  3.269452e-08  ...  0.028172  0.000623   
33  5.137857e-08  9.548780e-09  9.548780e-09  ...  0.006866  0.000108   
34  2.046956e-08  1.915463e-08  1.915463e-08  ...  0.014686  0.000216   
35  1.596396e-08  2.268898e-08  2.268898e-08  ...  0.022650  0.000256   
36  2.620009e-08  1.407441e-08  1.407441e-08  ...  0.011137  0.000159   
37  1.795991e-08  6.524601e-09  6.524601e-09  ...  0.003671  0.000073   
38  1.809025e-08  2.632373e-08  2.632373e-08  ...  0.060619  0.000296   
39  1.283586e-08  2.751992e-08  2.751992e-08  ...  0.017383  0.000310   
40  3.449979e-08  4.246660e-09  4.246660e-09  ...  0.004435  0.000048   
41  3.226594e-06  3.024027e-08  3.024027e-08  ...  0.044898  0.000341   
42  4.738518e-08  1.016696e-08  1.016696e-08  ...  0.011024  0.000115   
43  4.527317e-04  6.834258e-09  6.834258e-09  ...  0.005395  0.000077   
44  7.089076e-09  2.764847e-08  2.764847e-08  ...  0.021775  0.000311   
45  9.390308e-09  2.288793e-08  2.288793e-08  ...  0.021510  0.000258   
46  1.188359e-05  1.402732e-08  1.402732e-08  ...  0.020366  0.000158   
47  6.389264e-08  6.222844e-09  6.222844e-09  ...  0.004587  0.000070   
48  3.726048e-10  3.114976e-08  3.114976e-08  ...  0.040424  0.000351   
49  8.473119e-09  3.144257e-08  3.144257e-08  ...  0.043209  0.000354   
50  4.998116e-09  3.352802e-08  3.352802e-08  ...  0.052002  0.000378   
51  1.364949e-08  2.725022e-08  2.725022e-08  ...  0.026850  0.000307   
52  6.702715e-08  3.811868e-09  3.811868e-09  ...  0.002125  0.000043   
53  1.477752e-08  2.500337e-08  2.500337e-08  ...  0.033812  0.000282   
54  6.771162e-08  4.140914e-09  4.140914e-09  ...  0.003063  0.000055   
55  1.841679e-02  4.832648e-10  4.832648e-10  ...  0.000407  0.000005   
56  2.393092e-09  3.050337e-08  3.050337e-08  ...  0.017079  0.000351   
57  3.318478e-08  1.539335e-08  1.539335e-08  ...  0.008785  0.000173   

          23        24        25        26        27        28        29  \
id                                                                         
1   0.115699  0.005866  0.021850  0.000031  0.077242  0.001095  0.006526   
2   0.114341  0.003410  0.015195  0.000022  0.021868  0.000564  0.001981   
3   0.024684  0.023416  0.049637  0.000202  0.006253  0.007383  0.012547   
4   0.176399  0.003847  0.020421  0.000027  0.020782  0.000413  0.002940   
5   0.121270  0.014276  0.274932  0.000161  0.002668  0.000967  0.002729   
6   0.000052  0.043116  0.107266  0.000206  0.000011  0.006256  0.011167   
7   0.000078  0.019805  0.095237  0.000212  0.000025  0.005422  0.022694   
8   0.000206  0.078311  0.129290  0.000390  0.000009  0.006967  0.010040   
9   0.018965  0.013625  0.092816  0.000177  0.003089  0.002289  0.014736   
10  0.035855  0.009666  0.019314  0.000052  0.002990  0.000561  0.000703   
11  0.092015  0.019263  0.108094  0.000114  0.007596  0.002075  0.005840   
12  0.051786  0.059223  0.374352  0.000248  0.000264  0.001192  0.001423   
13  0.013124  0.018640  0.052865  0.000210  0.001067  0.003577  0.008987   
14  0.181400  0.066251  0.040491  0.000096  0.013693  0.007532  0.004746   
15  0.000563  0.027242  0.081673  0.000192  0.000053  0.002905  0.005808   
16  0.036754  0.055762  0.113471  0.000143  0.001202  0.001196  0.004210   
17  0.004093  0.008305  0.012349  0.000161  0.009967  0.006974  0.068220   
18  0.031099  0.013915  0.022385  0.000142  0.006749  0.004953  0.012171   
19  0.039630  0.060017  0.046816  0.000155  0.005411  0.009550  0.007900   
20  0.095128  0.046829  0.029166  0.000301  0.013124  0.022892  0.022115   
21  0.057092  0.021791  0.063754  0.000095  0.005012  0.001616  0.003392   
22  0.003974  0.021571  0.041587  0.000215  0.002399  0.004103  0.008146   
23  0.015388  0.009643  0.019771  0.000162  0.003743  0.001768  0.005251   
24  0.027928  0.001698  0.004029  0.000032  0.003765  0.000173  0.000173   
25  0.014254  0.033132  0.029596  0.000164  0.003736  0.006643  0.002734   
26  0.082215  0.031768  0.030960  0.000373  0.011697  0.001897  0.001897   
27  0.078969  0.010545  0.020511  0.000042  0.002120  0.000443  0.000226   
28  0.000095  0.013312  0.015522  0.000228  0.000010  0.001309  0.001398   
29  0.042854  0.018284  0.065101  0.000093  0.002069  0.000779  0.001065   
30  0.054574  0.052398  0.122136  0.000180  0.004532  0.005387  0.012916   
31  0.219361  0.016084  0.041382  0.000051  0.012882  0.002229  0.002363   
32  0.000847  0.114151  0.153417  0.007096  0.000136  0.007762  0.010439   
33  0.201573  0.036510  0.035536  0.000062  0.012494  0.003009  0.002928   
34  0.052654  0.036538  0.053567  0.000124  0.002378  0.004169  0.005004   
35  0.041752  0.047621  0.046893  0.000147  0.011016  0.005620  0.002706   
36  0.078388  0.029082  0.049013  0.000091  0.007908  0.004810  0.004007   
37  0.006714  0.002602  0.003479  0.000042  0.000689  0.000226  0.000293   
38  0.093509  0.024950  0.162515  0.000171  0.003899  0.002351  0.008503   
39  0.055119  0.113381  0.161825  0.000179  0.006309  0.002483  0.004523   
40  0.012825  0.001876  0.003391  0.000028  0.001379  0.000311  0.000700   
41  0.005666  0.024614  0.043943  0.000203  0.003246  0.008020  0.014145   
42  0.110739  0.015094  0.044887  0.000066  0.023176  0.002026  0.007717   
43  0.114799  0.004912  0.021888  0.000044  0.028936  0.000689  0.007027   
44  0.019854  0.057218  0.108907  0.000179  0.002327  0.006161  0.009630   
45  0.028069  0.064359  0.057022  0.000257  0.005433  0.007857  0.009324   
46  0.076146  0.016152  0.056417  0.000127  0.007826  0.001007  0.002357   
47  0.323062  0.012990  0.052856  0.000040  0.006125  0.000402  0.000588   
48  0.001728  0.012689  0.107462  0.000202  0.000013  0.002687  0.050811   
49  0.012655  0.025667  0.059400  0.000204  0.004836  0.009386  0.031794   
50  0.012840  0.029417  0.052854  0.000244  0.001221  0.007420  0.012814   
51  0.043987  0.026819  0.034925  0.000177  0.003715  0.004418  0.006425   
52  0.025983  0.001793  0.003307  0.000025  0.003268  0.000132  0.000700   
53  0.015227  0.032331  0.037326  0.000162  0.009199  0.014358  0.015661   
54  0.230042  0.010873  0.017122  0.000043  0.016006  0.000788  0.000843   
55  0.169513  0.000705  0.002079  0.000003  0.054992  0.000017  0.000154   
56  0.008593  0.124741  0.009902  0.000342  0.000587  0.015458  0.001078   
57  0.012217  0.006500  0.007919  0.000100  0.001217  0.000533  0.000533   

              30  
id                
1   3.311463e-06  
2   2.401088e-06  
3   1.941043e-05  
4   2.960754e-06  
5   1.737119e-05  
6   2.797145e-05  
7   2.226348e-05  
8   3.745849e-05  
9   1.909121e-05  
10  5.600064e-06  
11  1.234124e-05  
12  2.253246e-05  
13  2.260124e-05  
14  1.030686e-05  
15  2.074047e-05  
16  1.545478e-05  
17  1.736444e-05  
18  1.535626e-05  
19  1.673026e-05  
20  1.152258e-05  
21  1.022368e-05  
22  2.313905e-05  
23  1.727819e-05  
24  3.485431e-06  
25  1.763800e-05  
26  1.113115e-05  
27  4.564918e-06  
28  2.462878e-05  
29  9.977184e-06  
30  1.756400e-05  
31  5.445081e-06  
32  9.152766e-04  
33  6.680830e-06  
34  1.340159e-05  
35  1.587440e-05  
36  9.847201e-06  
37  4.564955e-06  
38  1.841747e-05  
39  2.742263e-04  
40  2.971187e-06  
41  2.115769e-05  
42  7.113344e-06  
43  4.781607e-06  
44  1.934433e-05  
45  7.035552e-05  
46  9.814252e-06  
47  4.353830e-06  
48  2.179402e-05  
49  2.199888e-05  
50  2.345797e-05  
51  1.906569e-05  
52  2.666984e-06  
53  1.749368e-05  
54  1.075735e-05  
55  3.381175e-07  
56  4.295529e-05  
57  1.077000e-05  

[57 rows x 31 columns]
\end{verbatim}

And save:

\begin{Shaded}
\begin{Highlighting}[]
\NormalTok{t1\_train\_metric }\OperatorTok{=}\NormalTok{ metrics.Task1Metric(train\_qrels.set\_index(}\StringTok{\textquotesingle{}id\textquotesingle{}}\NormalTok{), page\_kia, train\_qtarget)}
\NormalTok{binpickle.dump(t1\_train\_metric, }\StringTok{\textquotesingle{}task1{-}train{-}metric.bpk\textquotesingle{}}\NormalTok{, codec}\OperatorTok{=}\NormalTok{codec)}
\end{Highlighting}
\end{Shaded}

\begin{verbatim}
INFO:binpickle.write:pickled 1493808204 bytes with 5 buffers
\end{verbatim}

Do the same for eval:

\begin{Shaded}
\begin{Highlighting}[]
\NormalTok{eval\_qtarget }\OperatorTok{=}\NormalTok{ eval\_qrels.groupby(}\StringTok{\textquotesingle{}id\textquotesingle{}}\NormalTok{).}\BuiltInTok{apply}\NormalTok{(query\_xalign)}
\NormalTok{t1\_eval\_metric }\OperatorTok{=}\NormalTok{ metrics.Task1Metric(eval\_qrels.set\_index(}\StringTok{\textquotesingle{}id\textquotesingle{}}\NormalTok{), page\_kia, eval\_qtarget)}
\NormalTok{binpickle.dump(t1\_eval\_metric, }\StringTok{\textquotesingle{}task1{-}eval{-}metric.bpk\textquotesingle{}}\NormalTok{, codec}\OperatorTok{=}\NormalTok{codec)}
\end{Highlighting}
\end{Shaded}

\begin{verbatim}
INFO:binpickle.write:pickled 1493808200 bytes with 5 buffers
\end{verbatim}

\hypertarget{task-2-metric-preparation}{%
\subsection{Task 2 Metric Preparation}\label{task-2-metric-preparation}}

Task 2 requires some different preparation.

We're going to start by computing work-needed information:

\begin{Shaded}
\begin{Highlighting}[]
\NormalTok{page\_work }\OperatorTok{=}\NormalTok{ pages.set\_index(}\StringTok{\textquotesingle{}page\_id\textquotesingle{}}\NormalTok{).quality\_score\_disc.astype(pd.CategoricalDtype(ordered}\OperatorTok{=}\VariableTok{True}\NormalTok{))}
\NormalTok{page\_work }\OperatorTok{=}\NormalTok{ page\_work.cat.reorder\_categories(work\_order)}
\NormalTok{page\_work.name }\OperatorTok{=} \StringTok{\textquotesingle{}quality\textquotesingle{}}
\end{Highlighting}
\end{Shaded}

\hypertarget{work-and-target-exposure}{%
\subsubsection{Work and Target
Exposure}\label{work-and-target-exposure}}

The first thing we need to do to prepare the metric is to compute the
work-needed for each topic's pages, and use that to compute the target
exposure for each (relevant) page in the topic.

This is because an ideal ranking orders relevant documents in decreasing
order of work needed, followed by irrelevant documents. All relevant
documents at a given work level should receive the same expected
exposure.

First, look up the work for each query page ('query page work', or qpw):

\begin{Shaded}
\begin{Highlighting}[]
\NormalTok{qpw }\OperatorTok{=}\NormalTok{ qrels.join(page\_work, on}\OperatorTok{=}\StringTok{\textquotesingle{}page\_id\textquotesingle{}}\NormalTok{)}
\NormalTok{qpw}
\end{Highlighting}
\end{Shaded}

\begin{verbatim}
          id   page_id quality
0          1       572       C
1          1       627      FA
2          1       903       C
3          1      1193       B
4          1      1542      GA
...      ...       ...     ...
2199072  150  63656179   Start
2199073  150  63807245     NaN
2199074  150  64614938       C
2199075  150  64716982       C
2199076  150  65355704       C

[2199077 rows x 3 columns]
\end{verbatim}

And now use that to compute the number of documents at each work level:

\begin{Shaded}
\begin{Highlighting}[]
\NormalTok{qwork }\OperatorTok{=}\NormalTok{ qpw.groupby([}\StringTok{\textquotesingle{}id\textquotesingle{}}\NormalTok{, }\StringTok{\textquotesingle{}quality\textquotesingle{}}\NormalTok{])[}\StringTok{\textquotesingle{}page\_id\textquotesingle{}}\NormalTok{].count()}
\NormalTok{qwork}
\end{Highlighting}
\end{Shaded}

\begin{verbatim}
id   quality
1    Stub       1527
     Start      2822
     C          1603
     B           610
     GA          240
                ... 
150  Start       138
     C           127
     B            35
     GA           16
     FA            8
Name: page_id, Length: 636, dtype: int64
\end{verbatim}

Now we need to convert this into target exposure levels. This function
will, given a series of counts for each work level, compute the expected
exposure a page at that work level should receive.

\begin{Shaded}
\begin{Highlighting}[]
\KeywordTok{def}\NormalTok{ qw\_tgt\_exposure(qw\_counts: pd.Series) }\OperatorTok{{-}\textgreater{}}\NormalTok{ pd.Series:}
    \ControlFlowTok{if} \StringTok{\textquotesingle{}id\textquotesingle{}} \OperatorTok{==}\NormalTok{ qw\_counts.index.names[}\DecValTok{0}\NormalTok{]:}
\NormalTok{        qw\_counts }\OperatorTok{=}\NormalTok{ qw\_counts.reset\_index(level}\OperatorTok{=}\StringTok{\textquotesingle{}id\textquotesingle{}}\NormalTok{, drop}\OperatorTok{=}\VariableTok{True}\NormalTok{)}
\NormalTok{    qwc }\OperatorTok{=}\NormalTok{ qw\_counts.reindex(work\_order, fill\_value}\OperatorTok{=}\DecValTok{0}\NormalTok{).astype(}\StringTok{\textquotesingle{}i4\textquotesingle{}}\NormalTok{)}
\NormalTok{    tot }\OperatorTok{=} \BuiltInTok{int}\NormalTok{(qwc.}\BuiltInTok{sum}\NormalTok{())}
\NormalTok{    da }\OperatorTok{=}\NormalTok{ metrics.discount(tot)}
\NormalTok{    qwp }\OperatorTok{=}\NormalTok{ qwc.shift(}\DecValTok{1}\NormalTok{, fill\_value}\OperatorTok{=}\DecValTok{0}\NormalTok{)}
\NormalTok{    qwc\_s }\OperatorTok{=}\NormalTok{ qwc.cumsum()}
\NormalTok{    qwp\_s }\OperatorTok{=}\NormalTok{ qwp.cumsum()}
\NormalTok{    res }\OperatorTok{=}\NormalTok{ pd.Series(}
\NormalTok{        [np.mean(da[s:e]) }\ControlFlowTok{for}\NormalTok{ (s, e) }\KeywordTok{in} \BuiltInTok{zip}\NormalTok{(qwp\_s, qwc\_s)],}
\NormalTok{        index}\OperatorTok{=}\NormalTok{qwc.index}
\NormalTok{    )}
    \ControlFlowTok{return}\NormalTok{ res}
\end{Highlighting}
\end{Shaded}

We'll then apply this to each topic, to determine the per-topic target
exposures:

\begin{Shaded}
\begin{Highlighting}[]
\NormalTok{qw\_pp\_target }\OperatorTok{=}\NormalTok{ qwork.groupby(}\StringTok{\textquotesingle{}id\textquotesingle{}}\NormalTok{).}\BuiltInTok{apply}\NormalTok{(qw\_tgt\_exposure)}
\NormalTok{qw\_pp\_target.name }\OperatorTok{=} \StringTok{\textquotesingle{}tgt\_exposure\textquotesingle{}}
\NormalTok{qw\_pp\_target}
\end{Highlighting}
\end{Shaded}

\begin{verbatim}
C:\Users\michaelekstrand\Miniconda3\envs\wptrec\lib\site-packages\numpy\core\fromnumeric.py:3440: RuntimeWarning: Mean of empty slice.
  return _methods._mean(a, axis=axis, dtype=dtype,
C:\Users\michaelekstrand\Miniconda3\envs\wptrec\lib\site-packages\numpy\core\_methods.py:189: RuntimeWarning: invalid value encountered in true_divide
  ret = ret.dtype.type(ret / rcount)
\end{verbatim}

\begin{verbatim}
id   quality
1    Stub       0.114738
     Start      0.087373
     C          0.081146
     B          0.079298
     GA         0.078702
                  ...   
150  Start      0.154202
     C          0.127359
     B          0.120441
     GA         0.118827
     FA         0.118126
Name: tgt_exposure, Length: 636, dtype: float32
\end{verbatim}

We can now merge the relevant document work categories with this
exposure, to compute the target exposure for each relevant document:

\begin{Shaded}
\begin{Highlighting}[]
\NormalTok{qp\_exp }\OperatorTok{=}\NormalTok{ qpw.join(qw\_pp\_target, on}\OperatorTok{=}\NormalTok{[}\StringTok{\textquotesingle{}id\textquotesingle{}}\NormalTok{, }\StringTok{\textquotesingle{}quality\textquotesingle{}}\NormalTok{])}
\NormalTok{qp\_exp }\OperatorTok{=}\NormalTok{ qp\_exp.set\_index([}\StringTok{\textquotesingle{}id\textquotesingle{}}\NormalTok{, }\StringTok{\textquotesingle{}page\_id\textquotesingle{}}\NormalTok{])[}\StringTok{\textquotesingle{}tgt\_exposure\textquotesingle{}}\NormalTok{]}
\NormalTok{qp\_exp.index.names }\OperatorTok{=}\NormalTok{ [}\StringTok{\textquotesingle{}q\_id\textquotesingle{}}\NormalTok{, }\StringTok{\textquotesingle{}page\_id\textquotesingle{}}\NormalTok{]}
\NormalTok{qp\_exp}
\end{Highlighting}
\end{Shaded}

\begin{verbatim}
q_id  page_id 
1     572         0.081146
      627         0.078438
      903         0.081146
      1193        0.079298
      1542        0.078702
                    ...   
150   63656179    0.154202
      63807245         NaN
      64614938    0.127359
      64716982    0.127359
      65355704    0.127359
Name: tgt_exposure, Length: 2199077, dtype: float32
\end{verbatim}

\hypertarget{geographic-alignment}{%
\subsubsection{Geographic Alignment}\label{geographic-alignment}}

Now that we've computed per-page target exposure, we're ready to set up
the geographic alignment vectors for computing the per-\emph{group}
expected exposure with geographic data.

We're going to start by getting the alignments for relevant documents
for each topic:

\begin{Shaded}
\begin{Highlighting}[]
\NormalTok{qp\_geo\_align }\OperatorTok{=}\NormalTok{ qrels.join(page\_geo\_align, on}\OperatorTok{=}\StringTok{\textquotesingle{}page\_id\textquotesingle{}}\NormalTok{).set\_index([}\StringTok{\textquotesingle{}id\textquotesingle{}}\NormalTok{, }\StringTok{\textquotesingle{}page\_id\textquotesingle{}}\NormalTok{])}
\NormalTok{qp\_geo\_align.index.names }\OperatorTok{=}\NormalTok{ [}\StringTok{\textquotesingle{}q\_id\textquotesingle{}}\NormalTok{, }\StringTok{\textquotesingle{}page\_id\textquotesingle{}}\NormalTok{]}
\NormalTok{qp\_geo\_align}
\end{Highlighting}
\end{Shaded}

\begin{verbatim}
               Unknown  Africa  Antarctica  Asia  Europe  \
q_id page_id                                               
1    572           1.0     0.0         0.0   0.0     0.0   
     627           1.0     0.0         0.0   0.0     0.0   
     903           1.0     0.0         0.0   0.0     0.0   
     1193          1.0     0.0         0.0   0.0     0.0   
     1542          1.0     0.0         0.0   0.0     0.0   
...                ...     ...         ...   ...     ...   
150  63656179      1.0     0.0         0.0   0.0     0.0   
     63807245      NaN     NaN         NaN   NaN     NaN   
     64614938      1.0     0.0         0.0   0.0     0.0   
     64716982      1.0     0.0         0.0   0.0     0.0   
     65355704      1.0     0.0         0.0   0.0     0.0   

               Latin America and the Caribbean  Northern America  Oceania  
q_id page_id                                                               
1    572                                   0.0               0.0      0.0  
     627                                   0.0               0.0      0.0  
     903                                   0.0               0.0      0.0  
     1193                                  0.0               0.0      0.0  
     1542                                  0.0               0.0      0.0  
...                                        ...               ...      ...  
150  63656179                              0.0               0.0      0.0  
     63807245                              NaN               NaN      NaN  
     64614938                              0.0               0.0      0.0  
     64716982                              0.0               0.0      0.0  
     65355704                              0.0               0.0      0.0  

[2199077 rows x 8 columns]
\end{verbatim}

Now we need to compute the per-query target exposures. This starst with
aligning our vectors:

\begin{Shaded}
\begin{Highlighting}[]
\NormalTok{qp\_geo\_exp, qp\_geo\_align }\OperatorTok{=}\NormalTok{ qp\_exp.align(qp\_geo\_align, fill\_value}\OperatorTok{=}\DecValTok{0}\NormalTok{)}
\end{Highlighting}
\end{Shaded}

And now we can multiply the exposure vector by the alignment vector, and
summing by topic - this is equivalent to the matrix-vector
multiplication on a topic-by-topic basis.

\begin{Shaded}
\begin{Highlighting}[]
\NormalTok{qp\_aexp }\OperatorTok{=}\NormalTok{ qp\_geo\_align.multiply(qp\_geo\_exp, axis}\OperatorTok{=}\DecValTok{0}\NormalTok{)}
\NormalTok{q\_geo\_align }\OperatorTok{=}\NormalTok{ qp\_aexp.groupby(}\StringTok{\textquotesingle{}q\_id\textquotesingle{}}\NormalTok{).}\BuiltInTok{sum}\NormalTok{()}
\end{Highlighting}
\end{Shaded}

Now things get a \emph{little} weird. We want to average the empirical
distribution with the world population to compute our fairness target.
However, we don't have empirical data on the distribution of articles
that do or do not have geographic alignments.

Therefore, we are going to average only the \emph{known-geography}
vector with the world population. This proceeds in N steps:

\begin{enumerate}
\tightlist
\item
  Normalize the known-geography matrix so its rows sum to 1.
\item
  Average each row with the world population.
\item
  De-normalize the known-geography matrix so it is in the original
  scale, but adjusted w/ world population
\item
  Normalize the \emph{entire} matrix so its rows sum to 1
\end{enumerate}

Let's go.

\begin{Shaded}
\begin{Highlighting}[]
\NormalTok{qg\_known }\OperatorTok{=}\NormalTok{ q\_geo\_align.drop(columns}\OperatorTok{=}\NormalTok{[}\StringTok{\textquotesingle{}Unknown\textquotesingle{}}\NormalTok{])}
\end{Highlighting}
\end{Shaded}

Normalize (adding a small value to avoid division by zero - affected
entries will have a zero numerator anyway):

\begin{Shaded}
\begin{Highlighting}[]
\NormalTok{qg\_ksums }\OperatorTok{=}\NormalTok{ qg\_known.}\BuiltInTok{sum}\NormalTok{(axis}\OperatorTok{=}\DecValTok{1}\NormalTok{)}
\NormalTok{qg\_kd }\OperatorTok{=}\NormalTok{ qg\_known.divide(np.maximum(qg\_ksums, }\FloatTok{1.0e{-}6}\NormalTok{), axis}\OperatorTok{=}\DecValTok{0}\NormalTok{)}
\end{Highlighting}
\end{Shaded}

Average:

\begin{Shaded}
\begin{Highlighting}[]
\NormalTok{qg\_kd }\OperatorTok{=}\NormalTok{ (qg\_kd }\OperatorTok{+}\NormalTok{ world\_pop) }\OperatorTok{*} \FloatTok{0.5}
\end{Highlighting}
\end{Shaded}

De-normalize:

\begin{Shaded}
\begin{Highlighting}[]
\NormalTok{qg\_known }\OperatorTok{=}\NormalTok{ qg\_kd.multiply(qg\_ksums, axis}\OperatorTok{=}\DecValTok{0}\NormalTok{)}
\end{Highlighting}
\end{Shaded}

Recombine with the Unknown column:

\begin{Shaded}
\begin{Highlighting}[]
\NormalTok{q\_geo\_tgt }\OperatorTok{=}\NormalTok{ q\_geo\_align[[}\StringTok{\textquotesingle{}Unknown\textquotesingle{}}\NormalTok{]].join(qg\_known)}
\end{Highlighting}
\end{Shaded}

Normalize targets:

\begin{Shaded}
\begin{Highlighting}[]
\NormalTok{q\_geo\_tgt }\OperatorTok{=}\NormalTok{ q\_geo\_tgt.divide(q\_geo\_tgt.}\BuiltInTok{sum}\NormalTok{(axis}\OperatorTok{=}\DecValTok{1}\NormalTok{), axis}\OperatorTok{=}\DecValTok{0}\NormalTok{)}
\NormalTok{q\_geo\_tgt}
\end{Highlighting}
\end{Shaded}

\begin{verbatim}
       Unknown    Africa    Antarctica      Asia    Europe  \
q_id                                                         
1     0.575338  0.043635  3.278897e-08  0.153851  0.098450   
2     0.173889  0.069608  6.378567e-08  0.294269  0.280798   
3     0.234897  0.101882  5.907510e-08  0.278161  0.215027   
4     0.312664  0.076008  8.262075e-05  0.246140  0.145192   
5     0.182143  0.063760  6.314834e-08  0.273795  0.046710   
...        ...       ...           ...       ...       ...   
146   0.292441  0.090378  5.463208e-08  0.299627  0.067556   
147   0.434276  0.060053  4.368069e-08  0.195520  0.130625   
148   0.637050  0.033542  2.802409e-08  0.233693  0.045680   
149   0.370828  0.061724  4.857964e-08  0.243518  0.172170   
150   0.414091  0.062031  4.523918e-08  0.208319  0.131270   

      Latin America and the Caribbean  Northern America   Oceania  
q_id                                                               
1                            0.025042          0.065388  0.038296  
2                            0.046323          0.115193  0.019920  
3                            0.071196          0.077784  0.021053  
4                            0.058319          0.143947  0.017648  
5                            0.061549          0.366345  0.005697  
...                               ...               ...       ...  
146                          0.045686          0.178497  0.025815  
147                          0.061604          0.091005  0.026916  
148                          0.018613          0.025322  0.006099  
149                          0.040886          0.073876  0.036999  
150                          0.042203          0.116868  0.025218  

[106 rows x 8 columns]
\end{verbatim}

This is our group exposure target distributions for each query, for the
geographic data. We're now ready to set up the matrix.

\begin{Shaded}
\begin{Highlighting}[]
\NormalTok{train\_geo\_qtgt }\OperatorTok{=}\NormalTok{ q\_geo\_tgt.loc[train\_topics[}\StringTok{\textquotesingle{}id\textquotesingle{}}\NormalTok{]]}
\NormalTok{eval\_geo\_qtgt }\OperatorTok{=}\NormalTok{ q\_geo\_tgt.loc[eval\_topics[}\StringTok{\textquotesingle{}id\textquotesingle{}}\NormalTok{]]}
\end{Highlighting}
\end{Shaded}

\begin{Shaded}
\begin{Highlighting}[]
\NormalTok{t2\_train\_geo\_metric }\OperatorTok{=}\NormalTok{ metrics.Task2Metric(train\_qrels.set\_index(}\StringTok{\textquotesingle{}id\textquotesingle{}}\NormalTok{), }
\NormalTok{                                          page\_geo\_align, page\_work, }
\NormalTok{                                          train\_geo\_qtgt)}
\NormalTok{binpickle.dump(t2\_train\_geo\_metric, }\StringTok{\textquotesingle{}task2{-}train{-}geo{-}metric.bpk\textquotesingle{}}\NormalTok{, codec}\OperatorTok{=}\NormalTok{codec)}
\end{Highlighting}
\end{Shaded}

\begin{verbatim}
INFO:binpickle.write:pickled 2018 bytes with 9 buffers
\end{verbatim}

\begin{Shaded}
\begin{Highlighting}[]
\NormalTok{t2\_eval\_geo\_metric }\OperatorTok{=}\NormalTok{ metrics.Task2Metric(eval\_qrels.set\_index(}\StringTok{\textquotesingle{}id\textquotesingle{}}\NormalTok{), }
\NormalTok{                                         page\_geo\_align, page\_work, }
\NormalTok{                                         eval\_geo\_qtgt)}
\NormalTok{binpickle.dump(t2\_eval\_geo\_metric, }\StringTok{\textquotesingle{}task2{-}eval{-}geo{-}metric.bpk\textquotesingle{}}\NormalTok{, codec}\OperatorTok{=}\NormalTok{codec)}
\end{Highlighting}
\end{Shaded}

\begin{verbatim}
INFO:binpickle.write:pickled 2014 bytes with 9 buffers
\end{verbatim}

\hypertarget{intersectional-alignment}{%
\subsubsection{Intersectional
Alignment}\label{intersectional-alignment}}

Now we need to compute the intersectional targets for Task 2. We're
going to take a slightly different approach here, based on the
intersectional logic for Task 1, because we've come up with better ways
to write the code, but the effect is the same: only known aspects are
averaged.

We'll write a function very similar to the one for Task 1:

\begin{Shaded}
\begin{Highlighting}[]
\KeywordTok{def}\NormalTok{ query\_xideal(qdf, ravel}\OperatorTok{=}\VariableTok{True}\NormalTok{):}
\NormalTok{    pages }\OperatorTok{=}\NormalTok{ qdf[}\StringTok{\textquotesingle{}page\_id\textquotesingle{}}\NormalTok{]}
\NormalTok{    pages }\OperatorTok{=}\NormalTok{ pages[pages.isin(page\_xalign.indexes[}\StringTok{\textquotesingle{}page\textquotesingle{}}\NormalTok{])]}
\NormalTok{    q\_xa }\OperatorTok{=}\NormalTok{ page\_xalign.loc[pages.values, :, :]}
    
    \CommentTok{\# now we need to get the exposure for the pages, and multiply}
\NormalTok{    p\_exp }\OperatorTok{=}\NormalTok{ qp\_exp.loc[qdf.name]}
    \ControlFlowTok{assert}\NormalTok{ p\_exp.index.is\_unique}
\NormalTok{    p\_exp }\OperatorTok{=}\NormalTok{ xr.DataArray(p\_exp, dims}\OperatorTok{=}\NormalTok{[}\StringTok{\textquotesingle{}page\textquotesingle{}}\NormalTok{])}
    
    \CommentTok{\# and we multiply!}
\NormalTok{    q\_xa }\OperatorTok{=}\NormalTok{ q\_xa }\OperatorTok{*}\NormalTok{ p\_exp}

    \CommentTok{\# normalize into a matrix (this time we don\textquotesingle{}t clear)}
\NormalTok{    q\_am }\OperatorTok{=}\NormalTok{ q\_xa.}\BuiltInTok{sum}\NormalTok{(axis}\OperatorTok{=}\DecValTok{0}\NormalTok{)}
\NormalTok{    q\_am }\OperatorTok{=}\NormalTok{ q\_am }\OperatorTok{/}\NormalTok{ q\_am.}\BuiltInTok{sum}\NormalTok{()}
    
    \CommentTok{\# compute fractions in each section {-} combined with q\_am[0,0], this should be about 1}
\NormalTok{    q\_fk\_all }\OperatorTok{=}\NormalTok{ q\_am[}\DecValTok{1}\NormalTok{:, }\DecValTok{1}\NormalTok{:].}\BuiltInTok{sum}\NormalTok{()}
\NormalTok{    q\_fk\_geo }\OperatorTok{=}\NormalTok{ q\_am[}\DecValTok{1}\NormalTok{:, :}\DecValTok{1}\NormalTok{].}\BuiltInTok{sum}\NormalTok{()}
\NormalTok{    q\_fk\_gen }\OperatorTok{=}\NormalTok{ q\_am[:}\DecValTok{1}\NormalTok{, }\DecValTok{1}\NormalTok{:].}\BuiltInTok{sum}\NormalTok{()}
    
    \CommentTok{\# known average}
\NormalTok{    q\_am[}\DecValTok{1}\NormalTok{:, }\DecValTok{1}\NormalTok{:] }\OperatorTok{*=} \FloatTok{0.5}
\NormalTok{    q\_am[}\DecValTok{1}\NormalTok{:, }\DecValTok{1}\NormalTok{:] }\OperatorTok{+=}\NormalTok{ int\_tgt }\OperatorTok{*} \FloatTok{0.5} \OperatorTok{*}\NormalTok{ q\_fk\_all}
    
    \CommentTok{\# known{-}geo average}
\NormalTok{    q\_am[}\DecValTok{1}\NormalTok{:, :}\DecValTok{1}\NormalTok{] }\OperatorTok{*=} \FloatTok{0.5}
\NormalTok{    q\_am[}\DecValTok{1}\NormalTok{:, :}\DecValTok{1}\NormalTok{] }\OperatorTok{+=}\NormalTok{ geo\_tgt\_xa }\OperatorTok{*} \FloatTok{0.5} \OperatorTok{*}\NormalTok{ q\_fk\_geo}
    
    \CommentTok{\# known{-}gender average}
\NormalTok{    q\_am[:}\DecValTok{1}\NormalTok{, }\DecValTok{1}\NormalTok{:] }\OperatorTok{*=} \FloatTok{0.5}
\NormalTok{    q\_am[:}\DecValTok{1}\NormalTok{, }\DecValTok{1}\NormalTok{:] }\OperatorTok{+=}\NormalTok{ gender\_tgt\_xa }\OperatorTok{*} \FloatTok{0.5} \OperatorTok{*}\NormalTok{ q\_fk\_gen}
    
    \CommentTok{\# and return the result}
    \ControlFlowTok{if}\NormalTok{ ravel:}
        \ControlFlowTok{return}\NormalTok{ pd.Series(q\_am.values.ravel())}
    \ControlFlowTok{else}\NormalTok{:}
        \ControlFlowTok{return}\NormalTok{ q\_am}
\end{Highlighting}
\end{Shaded}

Test this function out:

\begin{Shaded}
\begin{Highlighting}[]
\NormalTok{query\_xideal(qdf, ravel}\OperatorTok{=}\VariableTok{False}\NormalTok{)}
\end{Highlighting}
\end{Shaded}

\begin{verbatim}
<xarray.DataArray (geography: 8, gender: 4)>
array([[5.40211229e-01, 1.22904624e-02, 2.26610467e-02, 1.75635724e-04],
       [3.80909493e-02, 2.90804953e-03, 2.59344827e-03, 4.25392005e-05],
       [2.85527900e-08, 2.09691080e-09, 2.09691080e-09, 4.23618344e-11],
       [1.34695670e-01, 8.88355123e-03, 1.01072347e-02, 1.64648516e-04],
       [8.71895859e-02, 2.97387866e-03, 8.25814408e-03, 2.84372324e-05],
       [2.16878846e-02, 1.67819032e-03, 1.65196427e-03, 2.36185304e-05],
       [5.32652519e-02, 2.51534798e-03, 9.59370956e-03, 1.36109402e-05],
       [3.48679417e-02, 4.71346052e-04, 2.95512391e-03, 1.46710935e-06]])
Coordinates:
  * geography  (geography) object 'Unknown' 'Africa' ... 'Oceania'
  * gender     (gender) object 'unknown' 'female' 'male' 'third'
\end{verbatim}

And let's go!

\begin{Shaded}
\begin{Highlighting}[]
\NormalTok{q\_xtgt }\OperatorTok{=}\NormalTok{ qrels.groupby(}\StringTok{\textquotesingle{}id\textquotesingle{}}\NormalTok{).progress\_apply(query\_xideal)}
\NormalTok{q\_xtgt}
\end{Highlighting}
\end{Shaded}

\begin{Shaded}
\begin{Highlighting}[]
\FunctionTok{\{}\DataTypeTok{"model\_id"}\FunctionTok{:}\StringTok{""}\FunctionTok{,}\DataTypeTok{"version\_major"}\FunctionTok{:}\DecValTok{2}\FunctionTok{,}\DataTypeTok{"version\_minor"}\FunctionTok{:}\DecValTok{0}\FunctionTok{\}}
\end{Highlighting}
\end{Shaded}

\begin{verbatim}
           0         1         2         3         4         5         6   \
id                                                                          
1    0.540211  0.012290  0.022661  0.000176  0.038091  0.002908  0.002593   
2    0.135109  0.010633  0.027958  0.000201  0.063400  0.003032  0.003115   
3    0.185923  0.018891  0.029817  0.000245  0.018607  0.033486  0.049321   
4    0.283620  0.008665  0.020234  0.000145  0.069568  0.003021  0.003361   
5    0.102865  0.021347  0.057531  0.000396  0.017768  0.022647  0.022888   
..        ...       ...       ...       ...       ...       ...       ...   
146  0.242344  0.017108  0.032738  0.000250  0.033631  0.031692  0.024843   
147  0.380085  0.025582  0.026067  0.001304  0.028472  0.017849  0.014999   
148  0.620663  0.005550  0.010755  0.000082  0.031143  0.001188  0.001188   
149  0.365415  0.002870  0.002516  0.000027  0.060143  0.000783  0.000783   
150  0.228180  0.057917  0.127065  0.000930  0.014522  0.021052  0.026136   

           7             8             9   ...        22        23        24  \
id                                         ...                                 
1    0.000043  2.855279e-08  2.096911e-09  ...  0.001652  0.000024  0.053265   
2    0.000059  5.789347e-08  2.916166e-09  ...  0.002811  0.000033  0.099662   
3    0.000471  1.300894e-08  2.280358e-08  ...  0.032680  0.000257  0.019061   
4    0.000059  8.261490e-05  2.894759e-09  ...  0.002071  0.000033  0.127457   
5    0.000458  1.758741e-08  2.255280e-08  ...  0.034300  0.000254  0.104245   
..        ...           ...           ...  ...       ...       ...       ...   
146  0.000212  3.349054e-08  1.046506e-08  ...  0.009259  0.000118  0.164611   
147  0.000207  2.317530e-08  1.019747e-08  ...  0.018809  0.000115  0.050914   
148  0.000024  2.563480e-08  1.182699e-09  ...  0.000659  0.000013  0.020264   
149  0.000016  4.700522e-08  7.793387e-10  ...  0.002786  0.000009  0.071839   
150  0.000320  1.332948e-08  1.579530e-08  ...  0.016691  0.000178  0.040721   

           25        26        27        28        29        30            31  
id                                                                             
1    0.002515  0.009594  0.000014  0.034868  0.000471  0.002955  1.467109e-06  
2    0.002826  0.012684  0.000019  0.017811  0.000468  0.001639  2.040304e-06  
3    0.018746  0.039812  0.000164  0.004908  0.005947  0.010179  1.595459e-05  
4    0.002601  0.013870  0.000019  0.015398  0.000279  0.001968  2.025326e-06  
5    0.012692  0.249255  0.000146  0.002342  0.000883  0.002456  1.577913e-05  
..        ...       ...       ...       ...       ...       ...           ...  
146  0.010455  0.003362  0.000068  0.013121  0.012324  0.000362  7.321910e-06  
147  0.022379  0.017457  0.000066  0.016122  0.005498  0.005220  7.134685e-06  
148  0.000380  0.004670  0.000008  0.003416  0.000041  0.002642  8.274784e-07  
149  0.000250  0.001781  0.000005  0.036944  0.000027  0.000027  5.452665e-07  
150  0.017971  0.058074  0.000103  0.009900  0.000547  0.013698  1.071915e-03  

[106 rows x 32 columns]
\end{verbatim}

\begin{Shaded}
\begin{Highlighting}[]
\NormalTok{train\_qtgt }\OperatorTok{=}\NormalTok{ q\_xtgt.loc[train\_topics[}\StringTok{\textquotesingle{}id\textquotesingle{}}\NormalTok{]]}
\NormalTok{eval\_qtgt }\OperatorTok{=}\NormalTok{ q\_xtgt.loc[eval\_topics[}\StringTok{\textquotesingle{}id\textquotesingle{}}\NormalTok{]]}
\end{Highlighting}
\end{Shaded}

\begin{Shaded}
\begin{Highlighting}[]
\NormalTok{t2\_train\_metric }\OperatorTok{=}\NormalTok{ metrics.Task2Metric(train\_qrels.set\_index(}\StringTok{\textquotesingle{}id\textquotesingle{}}\NormalTok{), }
\NormalTok{                                      page\_xa\_df, page\_work, }
\NormalTok{                                      train\_qtgt)}
\NormalTok{binpickle.dump(t2\_train\_metric, }\StringTok{\textquotesingle{}task2{-}train{-}metric.bpk\textquotesingle{}}\NormalTok{, codec}\OperatorTok{=}\NormalTok{codec)}
\end{Highlighting}
\end{Shaded}

\begin{verbatim}
INFO:binpickle.write:pickled 1879 bytes with 9 buffers
\end{verbatim}

\begin{Shaded}
\begin{Highlighting}[]
\NormalTok{t2\_eval\_metric }\OperatorTok{=}\NormalTok{ metrics.Task2Metric(eval\_qrels.set\_index(}\StringTok{\textquotesingle{}id\textquotesingle{}}\NormalTok{), }
\NormalTok{                                     page\_xa\_df, page\_work, }
\NormalTok{                                     eval\_qtgt)}
\NormalTok{binpickle.dump(t2\_eval\_metric, }\StringTok{\textquotesingle{}task2{-}eval{-}metric.bpk\textquotesingle{}}\NormalTok{, codec}\OperatorTok{=}\NormalTok{codec)}
\end{Highlighting}
\end{Shaded}

\begin{verbatim}
INFO:binpickle.write:pickled 1875 bytes with 9 buffers
\end{verbatim}

\begin{Shaded}
\begin{Highlighting}[]

\end{Highlighting}
\end{Shaded}